\documentstyle[draft]{mn}
\newcommand{\mnref}[1]{\hangindent=0.5in \hangafter=1 #1 \par}

\newcommand{\mn}{MNRAS}

\newcommand{\aj}{AJ}
\newcommand{\apj}{ApJ}
\newcommand{\apjs}{ApJS}

\title[The Cluster Galaxy Luminosity Function]
{The Edinburgh-Durham Southern Galaxy Catalogue -- \\ VIII:
The Cluster Galaxy Luminosity Function}
\author[S.L. Lumsden et al.]
{S.L. Lumsden$^{1}$, C.A. Collins$^{2}$, R.C. Nichol$^{3}$, V.R. Eke$^{1}$, 
and L. Guzzo$^{4}$\\
{}$^1$ {\em Anglo-Australian Observatory, PO Box 296, Epping, NSW 2121,
Australia} \\
\hspace{1cm} {Email -- sll@aaoepp.aao.gov.au, v.r.eke@durham.ac.uk}\\
{}$^2$ {\em Astrophysics Group, School of 
Electrical Engineering, Electronics and Physics, 
Liverpool John Moores University,} \\
\hspace{1cm} {Byrom Street, Liverpool L3 3AF} \\
\hspace{1cm} {Email -- cac@staru1.livjm.ac.uk}\\
{}$^3$ {\em Department of Physics, Carnegie Mellon University,
5000 Forbes Avenue, Pittsburgh, PA15213-3890, USA} \\
\hspace{1cm}{ Email -- nichol@astro.phys.cmu.edu}\\
{}$^4$ {\em Osservatorio Astronomico di Brera, Via Bianchi 46, I-22055 
Merate, Italy}  \\
\hspace{1cm} {Email -- guzzo@astmim.mi.astro.it}\\
}

\begin{document}
\label{firstpage}

\maketitle

\begin{abstract}
We have re-examined the nature of the cluster galaxy luminosity function using
the data from the Edinburgh-Durham Southern Galaxy Catalogue and the
Edinburgh-Milano Redshift Survey.  We derive a best fit
luminosity function
over the range $-18$
to $-21$ in $M(b_j)$, for a composite sample of 22 of the richer clusters that has
$M(b_j)^*=-20.16\pm0.02$ and $\alpha=-1.22\pm0.04$.
The dominant error in
these values results from the choice of background subtraction method.  From
extensive simulations we can show that when the LF is fitted over this narrow
range, it is difficult to discriminate against bright values of $M^*$ in the
single cluster fits, but that faint values provide a strong test of the
universality of the luminosity function.  We find that all the individual
cluster data are well fit by a Schechter function with $\alpha$ fixed at
$-1.25$, and that $\le$10\% of these have fitted values of $M^*$ that disagree
from the average at the 99\% confidence level.  
We further show that fitting only a
single parameter Schechter function to composite subsets of the data can give
erroneous results for the derived
$M^*$, as might be expected from the known tight correlation between
$M^*$ and $\alpha$.  By considering two parameter fits, the results of
Monte-Carlo simulations and direct two-sample $\chi^2$ tests we conclude that
there is only weak evidence for differences between the data when broken down
into subsets based on physical properties (Bautz-Morgan class, richness,
velocity dispersion):  from our simulations, only the evidence for a difference
between subsets based on velocity dispersion may in fact be
significant.  However, we find no evidence at all that a Schechter function
is not a good model for the intrinsic cluster luminosity function over this
absolute magnitude range.  Models that invoke strong evolution of galaxy
luminosity of {\em all}
 galaxies within clusters are inconsistent with our results.
\end{abstract}


\section{Introduction}
The properties of the cluster galaxy
luminosity function (LF) are still a matter of 
active debate. On the one hand considerable stress has been laid on a  single 
analytic expression for the LF, as 
determined by Schechter (1976), which it is argued holds true for all clusters 
and the field (e.g. Abell 1975). On the 
other hand, much work has been concerned to establish differences in the LF 
as a function of cluster morphology. Indeed, a universal 
LF is hard to explain theoretically. For example, according to the cannibalism 
line of argument (e.g. Hausman and Ostriker 1978), cD galaxies grow at the 
expense of other fainter galaxies - thus such a model predicts a deficit of 
bright galaxies in the most dynamically evolved clusters. The brightness of 
the 1st ranked 
cluster galaxy is thus related to the dynamical state of the cluster and 
the LF is expected to correlate with Bautz-Morgan (B-M) type, as this 
measures the contrast of the 1st ranked galaxy 
compared to the other bright galaxies. Such a deficit in bright galaxies is 
also expected if tidal stripping in the outer halos of galaxies, due to 
interactions, is taking place 
(Richstone 1975, 1976; Merritt 1983; Malumuth \& Richstone 1984). Only in the 
special case in which the luminosity function is determined during the 
formation period of the cluster, with little further dynamical evolution taking 
place,  are no correlations of the LF with other cluster 
properties expected (Merritt 1984, 1985).

The most popular approach taken to the study of the LF is to fit 
Schechter's (1976) parameterized form, given by
\[n(L) dL=n^{*}(L/L^{*})^{\alpha}exp(-L/L^{*}) d(L/L^{*}),\]
where $n^{*}$ is a normalization factor, $\alpha$ is the power-law slope at 
the faint end, and $L^{*}$ (or equivalently
$M^{*}$) is the characteristic luminosity 
(magnitude). Schechter(1976) determined the best-fit parameter values to be 
$M^{*}=-21.41$ (J band) and $\alpha=-1.25$. In 
subsequent studies there has 
been claim and counter claim regarding the properties of this function. In 
a study of 12 very rich clusters, Dressler 
(1978) concluded that $M^*$ does vary significantly from cluster to cluster 
and that it is correlated with the absolute magnitude of the first-ranked 
galaxy, in the sense consistent with the trend expected in the cannibalism 
model. Lugger (1986) studied a sample of 9 Abell clusters and concluded that
while the mean values of $M^{*}$ and $\alpha$ are in good agreement with those 
of Dressler(1978) and Schechter(1976), there was no evidence for a correlation 
of $M^{*}$ with cluster morphology, central density or magnitude of the 
first-ranked galaxy. The examination of a further 14 rich clusters by Colless
(1989), draws very much the same conclusions as Lugger(1986) - with no 
variation in the cluster LFs with B-M type or richness seen. Simulations 
suggest 
that variations of more than 0.4 mag in $M^{*}$ or 0.15 in $\alpha$ are ruled 
out by these data. Colless did find a marginally significant result 
such that the LF of the high velocity dispersion clusters has fainter $M^{*}$ 
compared with the low velocity cases.

There are potential criticisms of these previous studies: first, as pointed out
by Lugger(1986), the early studies such as that of Dressler (1978) did not use
a consistent cluster radius or limiting absolute magnitude in comparing
different cluster LFs.  Secondly, the best previous studies (those of Dressler
1978, Lugger 1986 and Colless 1989) confine their selection to Abell clusters,
and typically only the richer examples of these systems.  Since there are
potential systematic biases in the way in which Abell clusters were selected
(for example, cD dominated clusters are much easier to detect by eye), these
studies cannot necessarily be held to be a true reflection of the properties of
all clusters.  Last, previous studies have relied on global number counts
 to correct for the non-cluster background present, and have used counts often
taken in different passbands and from different surveys.  This can lead to
systematic errors due to differences in the calibrations of the data the LF
is derived from and the data the counts are taken from.

Here we have re-examined the LF afresh using a sample of 46 clusters,
selected automatically from digitized data. As a result this study has certain
advantages over previous work in the field. The clusters are selected from
photometrically calibrated digitized scans of photographic plates using an
automatic search algorithm. Hence our data should be more uniform than the
Abell catalogue. The sample covers a wide range in richness and many of the
clusters have well determined velocity dispersions, allowing us to test for
variations in the LF as a function of intrinsic cluster properties.  

\section{The cluster sample}
\subsection{The catalogue and the redshift survey}
The sample of clusters used in this paper
is drawn from the Edinburgh/Durham Cluster Catalogue
(hereafter EDCC: Lumsden et al.\ 1992), which in its turn is derived from the
Edinburgh/Durham Southern Galaxy Catalogue (hereafter EDSGC: eg
Heydon-Dumbleton et al.\ 1989).  The EDSGC consists of digitized COSMOS scans
of 60 UK Schmidt Telescope III-aJ survey plates centred on the south galactic
pole, and is nominally 95\% complete to $b_j=20$ with less than 10\%
contamination.  A cluster sample was derived from the galaxy data in the EDSGC
as outlined in Lumsden et al.  There are over 700 potential clusters in the
EDCC, and this sample is the first published machine based cluster sample.
The 
objective nature of the construction of the EDCC
implies that it does not suffer from the
systematic subjective biases that appear in the southern extension of the Abell
catalogue covering the same area (Abell, Corwin \& Olowin, 1989).  The EDCC is
complete within its selection criteria to a limiting magnitude for the tenth
ranked cluster member of $b_j=18.75$.

For a smaller sub-sample of the EDCC clusters, we obtained redshifts 
with the multi-slit spectrograph EFOSC on the ESO 3.6m and
with the
multi-fibre spectrograph Autofib at the 3.9m Anglo-Australian Telescope.
The nature of this
sample is discussed in Nichol et al.\ (1992) and Guzzo et al.\ (1992)
and the redshifts are published in
Collins et al.\ (1995). Approximately 10\% of those clusters observed proved to
have significant contamination from other groups along the line of sight at
significantly different redshifts.  We have excluded those clusters with
`projection effects' as defined in Collins et al.\ (1995) from our sample.  We
note that the remaining clusters do not have a well defined cut-off in cluster
richness (unlike the sample of Nichol et al), since we have included all the
clusters for which we actually obtained redshifts.

From this dataset, we have excluded those clusters where the cluster background
radius (as defined in section 3.3) overlaps the boundaries of the EDSGC for
ease of computation, and those clusters where their respective Abell radii
are entirely superposed,
since there is then no well defined cluster magnitude distribution.  We also
individually checked all the other clusters where the two clusters' Abell radii
overlapped since the deblending procedure described in Lumsden et al.\ can
occasionally cause the derived cluster centroid to move away from the `true'
position: we removed those clusters where the catalogue positions differed from
both the centre derived from a visual inspection of the plate and the centroid
of those galaxies for which we have redshifts by a significant fraction
($\sim$30\%) of the Abell radius.  For blended clusters that pass this test,
we assign galaxy membership to the individual clusters using the method
described in Lumsden et al.\ (1992).
Lastly, to be confident that we have removed
all clusters that may have contaminating projection effects, we choose to use
only those with a minimum of 4 cluster redshifts, and where the number of
galaxies with redshifts outside of the cluster is less than the actual number
of {\em bona-fide} cluster members.

The parameters of the remaining 46 clusters (as taken from Lumsden et al.\ 1992
and Collins et al.\ 1995) are given in Table 1.  For all of the clusters, we
have considered cluster members to be those galaxies within one Abell radius
(where we have taken $r_A=1.5h^{-1}$Mpc and we use the value for the Hubble
constant $h=1$ throughout
\footnote{Where $h=$H$_0/100$, and H$_0$
is the Hubble constant in units of km$^{-1}$s$^{-1}$Mpc$^{-1}$}).
Our sample
is the largest single database of clusters which has been used to study the LF
to date.

\subsection{Magnitude calibration}
The EDSGC was calibrated from photometric CCD sequences assuming that the plate
and CCD magnitudes were linearly related (Heydon-Dumbleton et al.\ 1989).
However, for bright objects whose profiles are either strongly peaked (such as
elliptical galaxies) or stellar, there is a regime in which the measured COSMOS
magnitude is less than the true magnitude, caused by lack of dynamic range
within the measuring machine itself.  This `saturation' occurs for elliptical
galaxies at $b_j\sim17$ (note that the plates themselves only saturate at much
brighter magnitudes).  Fainter than this magnitude, and for most late type
galaxies, the original calibration is correct, so that previously published
results from the EDSGC are unchanged by these corrections.  However, since most
clusters are dominated by elliptical galaxies at the bright end of the LF it is
vital that we correct for this effect or else we will induce errors in the
measured value of the `break' in the LF for the nearby clusters.

We have calculated a correction for this effect, by comparing the catalogued
EDSGC magnitude against the observed magnitude for 451 galaxies present in the
CCD sequences used to calibrate the EDSGC initially (Heydon-Dumbleton et al).
Since for faint magnitudes the slope should be linear, but at bright magnitudes
it should diverge from this, we used a quadratic fit to the data.  From this we
derive a fit with the form (see Fig.\ 1)
\[ b_j({\rm EDSGC}) -b_j({\rm CCD}) = 
      (1.06\pm0.26)\times10^{-2}b_j({\rm EDSGC})^2 \]
\vspace*{-3mm}
\[ \hspace*{0.8in}      -(0.46\pm0.09)b_j({\rm EDSGC}) + (4.96\pm0.90). \]
This fit should be reliable in the range $15<b_j<21$.  For most of this range,
the linear component present is actually the most important.  This slope has
been commented on previously by Heydon-Dumbleton (1989), and gives rise to the
observed slope in the fit at fainter magnitudes, and is a result of the
definition of measured photographic magnitude used in the EDSGC.  As can be
seen the effect is small (the overall effect on the composite LF described in
section 5.1 for example is 0.2 magnitudes).  However, for the purposes of the
LF, we include all terms, so that our magnitude scale is more closely matched
to the CCD based magnitude scale.

It should be noted that this correction is an average, and may not represent
every cluster absolutely.  Since the saturation effect is a function of the
photographic density, anything else that modifies this (such as vignetting,
emulsion flaws, plate fogging or emulsion desensitisation near the plate
corners) will consequently affect the amount of saturation present.  We do not
have sufficient CCD sequences to map these effects fully however, although we
note that the overall photographic intensity--magnitude relation has been
corrected for these effects (see Heydon-Dumbleton 1989).  These effects may 
in part give rise to the observed scatter in Fig.\ 1.  We expect the
magnitude of any error caused by using this average calibration to be small,
since the deviations from the fit are small.

\section{The cluster galaxy luminosity function}
\subsection{Form of the luminosity function}
As is common with studies of the cluster galaxy luminosity function
we use the standard Schechter function (Schechter 1976)
as the analytical model to our data.  
This function has the following form when expressed in terms of absolute
magnitude:
\[
{\rm n(M)dM} = kn^* \times \]
\vspace*{-3mm}
\begin{equation}\hspace*{1cm}\exp\{k(\alpha+1)({\rm M}^*-{\rm M}) - 
      \exp(k({\rm M}^*-{\rm M}))\}{\rm dM}, \end{equation}
where $k={\rm ln}10/2.5$. 
Here $M$ is the absolute magnitude, derived from the measured apparent
magnitude $m$ using,
\[ M = m -42.384 + 5\log h - 5\log z - K(z) -A(b_j),\]
where 
$K(z)=4.14z-0.44z^2$ is the K-correction suitable for the $b_j$ passband
for the mix of galaxies typically observed in low redshift clusters
(Ellis 1983), and $A(b_j)$ is the effective extinction
in this passband.  This latter quantity is derived using techniques described
in Nichol \& Collins (1993) for each cluster (see also appendix A).
In common with other studies of the luminosity function we assume that
$q_0=1$ in deriving the distance modulus used, and we have adopted $h=1$
for simplicity as well.
The LF normalisation $n^*$ is constrained
using the actual observed number of cluster galaxies, and we
carry out either a two parameter fit to 
$M^*$ (the absolute magnitude of the `break' in the LF) and $\alpha$ (the
exponent of the power-law slope at fainter magnitudes), or a one parameter
fit to $M^*$ with $\alpha$ fixed where required.

\subsection{The Composite Luminosity Function}
There are too few galaxies per bin in the individual LFs to fit both $M^*$ and
$\alpha$ simultaneously.  Since we wish to search for deviations from some
`mean' LF, it is valuable to consider coadding all the LF data into an average
or composite luminosity function, in a way that allows us to estimate both
parameters simultaneously.  The easiest manner in which to do this is to
consider the composite luminosity function defined by Colless (1989) to be
\begin{equation}
N_{cj} = \frac{R_c}{n_{clus,j}} \Sigma_{clus}\frac{N_{ij}}{R_i}, \end{equation}
where $R_i$ is the `richness' of the $i$th cluster,
$n_{clus,j}$ is the number of clusters that actually contribute 
to the $j$th bin of the composite LF, $N_{ij}$ is the background corrected
number of galaxies in the $j$th bin of the $i$th cluster's LF and
$R_c = \Sigma_{clus} R_i$.  
We use a different definition of $R_i$ to that of
Colless, since he used the total number of galaxies brighter than $M=-19$
and we use
the background corrected number of cluster galaxies between 
$M(b_j)=-19.5$ and $M(b_j)=-20.5$. 
This is because
our sample contains more distant clusters on average than the Colless
sample, and we
would otherwise discard a significant fraction of the data as being objects
where $M=-19$ was beyond our chosen completeness limit.  We also impose
an upper limit in absolute magnitude
when calculating the richness to avoid the problems with very
bright galaxies discussed above.  For typical values for the LF ($M^*=-20$ and
$\alpha=-1.25$), the relationship between our definition of richness and that
of Colless  is
$R_i$(Colless)$\sim2R_i$(here).

Similarly, the errors for the composite LF are calculated according to 
\begin{equation}
\delta N_{cj} = \frac{R_c}{n_{clus,j}} \left[
\Sigma_{clus} \left(\frac{\delta N_{ij}}{R_i}\right)^2\right]^{1/2},
\end{equation}
where the $\delta N_{ij}$ are derived according to the error given in Section
3.4.

\subsection{Background galaxies}

In order to derive the LF, we must be able to define the expected contribution
from background galaxies, $n_{back}$, accurately.  We have considered three
approaches to this problem.  First, we used the counts in an annulus, centred
on the cluster, with an inner radius of one degree and a width of one degree.
This fixed radius was adopted for simplicity --
using a fixed metric radius as opposed to a fixed angular radius gives
essentially the same results.
This method has the advantage that the correction is strictly local, and hence
the variation in field counts seen in different regions of the EDSGC is
correctly accounted for. 
For the other two methods we used the
the global counts as derived from
the whole of the EDSGC, and the average of the counts from the annuli from the
first method described above.  
The first of these is similar to the procedure used in previous
studies, but since our global counts actually come from the same data the LFs
themselves are constructed from, we remove the possibility of errors arising
from discrepancies in the calibration of the LF and the global counts.  
The second has the potential advantage that it excludes the rich clusters
from the counts and hence is a better measure of the `true' background. In
Fig.\ 2
we show the derived counts from the catalogue as a whole together with the
average of the counts derived from the annuli around the clusters.  There is
essentially little difference between these {\em on average}.  The difference
that does exist is at the level 
expected from 
the presence of real large-scale structure.  There is little sign that
excluding the rich clusters cores makes the background as calculated from
the annuli lower than the global counts.

Previous estimates of the background count for use in the LF have been
derived from either separate `eye-ball' galaxy catalogues (e.g.\ Dressler 1978,
Lugger 1986) or have been based on data from only very small areas (Colless
1989).  We note in passing that the slope of the number counts that we find is
larger than that assumed in previous studies of the LF.  This is however
consistent with Maddox et al.\ (1990), the only other measure of the counts
over significantly large areas of the sky that is derived using a measuring
machine.  The fit given by Colless (1989) appears to overcount the true
background by $\sim20$\%.

Lastly, we note that the derived number counts shown in Fig.\ 2 drop off more
rapidly than $m^{0.45}$ at $b_j>20$.  This disagrees with other observations
(eg the deep CCD counts presented in Metcalfe et al.\ 1991), and indicates that
the overall completeness limit of the EDSGC is $b_j\sim20$.  However,
inspection of the number counts on individual Schmidt plates also shows that
this limit varies from plate to plate.  We have allowed for a possible variable
completeness with plate in our work by examining the number counts on each
plate separately and assigning completeness limits relative to the 7 deepest
fields in the survey.  From this we find that the best plates are complete to
$b_j=21$ and the worst to $b_j=19.5$.  We do not however use data fainter than
$b_j=20$ from any field.  For those fields where the adopted completeness
limit is brighter than this, we do not use data fainter than that limit.
In calculating the background correction required
when using the global number counts we have interpolated the actual measured
data, rather than fitting a single power law.  The latter is a poor
representation to the actual data, since the true number counts `turn-over'
near $b_j\sim19$ (eg Metcalfe et al.).  This regime is important for our work
since for clusters with $z\sim0.1$, $b_j=20$ is equivalent to an absolute
magnitude of $\sim-18$.

We also have to derive suitable error estimates for these background counts.
There are two sources of error, which are present in all the methods given
above.  The first is simply the Poisson uncertainty in the counts and can be
represented simply as $\sqrt{n_{back}}$.  The second is systematic and
represents the presence of large-scale clustering within the sample. For the
global correction method, although we would assume that the slope of the
background number counts is constant over the survey region, the normalisation
of these counts changes with position.  Colless (1989) and Lugger (1986) follow
Geller \& Beers (1982) in assuming that the background is only known to 50\%.
For the background derived from the local annuli we can estimate the error
rather more accurately however. If we take all of these annuli, and find the
scatter about the mean of the number counts, then we can derive the systematic
error for this method. Since all the annuli have the same area
any number density deviations are purely due to the presence of large-scale
structure.  The derived variance is also shown on Figure 2.
Comparison of the expected contribution from Poisson error and the
total error observed indicates that the mean field-to-field variance is 25\%
for $b_j > 17$.  The largest excursions seen are $<$50\% over the same
magnitude range.  Therefore, using the 50\% variation in what follows, although
conservative, does account for all the observed variations, whereas using the
25\% mean variation is a better representation of the data as a whole but will
underestimate the error in the background for a few clusters.  We demonstrate
in section 5.1 the effect of using this smaller error on the field counts
compared to the value favoured by Geller \& Beers.  We note that Dressler
(1978) also found that a mean variation of 25\% was a good fit to his data, and
this result is compatible with the known angular clustering in the EDSGC
(Collins, Nichol \& Lumsden 1992).

\subsection{Fitting Methods}
We have considered only a $\chi^2$ squared fitting method in deriving the
parameters $M^*$ and $\alpha$.  If we let
$n_{clus}$ be the difference between the total counts in a bin and the expected
background in the bin (however the latter is calculated), so that
\[ n_{clus} = n_{tot} - n_{back},\]
then
we seek to minimise the function
\begin{equation} \chi^2 = \Sigma \frac{(n_{clus} -
n_{fit})^2}{\sigma(n_{fit})^2},\end{equation}
where,
\[n_{fit} = n(M)\Delta M + \frac{n''(M) \Delta M^3}{24.0} \]
and 
\[ \sigma(n_{fit}) = [\sigma(n_{tot})^2 + \sigma(n_{back})^2]^{1/2}.\]
Throughout we use a bin width, $\Delta M$, of 0.5 magnitudes.
The presence of the second derivative in the form of $n_{fit}$ takes account
of the finite bin width used, since the number counts change extremely rapidly
(Schechter 1976).
The first term in the error 
represents the Poissonian error in the total observed counts.  Therefore,
$\sigma(n_{tot})=n_{tot}^{1/2}$.
The second term
must take account of both Poissonian error and systematic errors in the
background as noted in section 3.3.  This has the form
$\sigma(n_{back}) = {\rm max}(n_{back}^{1/2},x n_{back})$, where
$x=0.5$ (except where explicitly noted that $x=0.25$).

Although for many of the clusters data exists with $M$ brighter than $-21$
and
fainter than $-18$, there are good reasons for not using such data in deriving
values of $M^*$ and $\alpha$.  First, for the fainter galaxies, the correction
for the background becomes more difficult since the background fraction per
bin rises steadily.  For any given cluster we impose two limits on this: that
any data with $m(b_j)>20$ be discarded, for reasons outlined in section 3.3,
and that any bin for which the number of background galaxies exceeds the
cluster galaxies also be discarded.  In practice, the former limit essentially
ensures the latter for faint galaxies in any event.  For bright galaxies there
are also problems we wish to avoid.  
It is desirable,
as has been noted previously by, e.g.\,
Lugger, to exclude the brightest cluster galaxy when deriving
the LF.  We automatically do this since we discard all galaxies
brighter than $M=-21$ (typically of order five galaxies in the richer
clusters). There are more specific reasons for doing this in our case 
as well.  First, the saturation correction is poorly
defined brighter than $b_j=15$, and hence the magnitudes of the galaxies
brighter than this are likely to have larger errors.  At $z=0.05$, this
magnitude corresponds to $M=-21.5$, and at $z=0.1$ $M=-23$.  Therefore, 
it is clear that galaxies brighter than $M=-21$ should not be used because
their photometry is suspect.  Secondly, 
COSMOS data is deblended where object mergers occur (cf the discussion in
Lumsden et al.\ 1992), but the accuracy of the
magnitudes of the deblended data is less reliable than for single isolated
objects.  The brightest galaxies in clusters are almost always blended
with fainter satellite systems in the EDSGC data, so again the photometry of
these objects is not reliable.
Discarding the brightest bins of the LF guards
against problems of this kind.  
It is possible that similar problems may also be present in previous
derivations of the LF, since the work described in the introduction also
relied extensively on photographic plate material.  
In any event we recommend that results which rely
on accurate magnitudes for the brightest galaxies in any LF derived from
photographic plates should be treated with caution.
We have also placed constraints on the values
of $M^*$ and $\alpha$ that the fitting routine can accept, to prevent any
possibility of the fitting process `running away.' These limits are that $M^*$
should lie between $-18$ and $-26$ and that $\alpha$ should lie between 0 and
$-2.5$.  Therefore quoted values for the fits that are near these limits are
indicative of a very shallow minimum in the fit, and hence that the reliability
of the fit is likely to be low (whatever the quoted formal probability that the
fit is good).  Lastly, we note that we sum bins to ensure that there are $\ge5$
galaxies per bin in the data being fitted, as is standard for $\chi^2$
minimisation techniques.

\section{Reliability of the Fitting Procedure}
Before we can consider the actual data,
we need some estimate of the reliability of
our fitting procedure and the true error on these fits.  To achieve this, we
conducted extensive Monte-Carlo simulations.  In particular, we carried out
three sets of simulations: first, to estimate the effect that errors in the
assumed background had on the fit parameters as a function of redshift;
secondly, to derive a measure of the true error distribution for a one
parameter fit to a model LF as a function of richness, for $R_i$ typical of the
single cluster values we found; lastly, to derive the error distribution for
the two parameter fits to the composite sets.  These latter two will then allow
us to estimate the true likelihood that observed departures from the composite
LF are in fact truly significant.  In carrying out these simulations, we have
assumed that $M^*=-20.2$ and $\alpha=-1.25$ for the model of the LF
(in accordance with the results derived in section 5.1).

In order to check the effect of the background subtraction alone, we initially
assumed that the form of the LF was known perfectly, and only the background
was allowed to vary.  We used the global background as derived above, and
rescaled it at random so that the variance was 25\% as was found to be the case
in practise in section 3.3. Thus after normal background correction, the
random LF would have either too many or too few background galaxies subtracted.
By deriving fit parameters for many such datasets we can derive the average
error induced in those parameters as a function of cluster richness and
redshift.  In general, for $z<0.05$, the effect of errors in the background on
the fit parameters is negligible, even for poor clusters.  At higher redshifts,
$M^*$ can be measured to an accuracy of 0.05 magnitudes if $R_i>30$
(ie the average of the measured $M^*$ from the simulations is different
by this amount from the assumed input $M^*$).  At
$z>0.05$ and $R_i<20$, the scatter in the derived value of $M^*$ becomes larger
($\sim0.5-1.0$ magnitudes).  The values derived here should be the dominant
source of error when the richness is very large (since then the LF is known
almost perfectly).  Since the composite luminosity functions given all have
effective values of the richness that are much greater than 100, it is clear
that the empirically derived errors for the effect of different background
corrections given in section 5.1 are in good agreement with the results of
these Monte-Carlo simulations.

More generally, we need to map out true estimates of the likely scatter in the
fits to $M^*$ when the LF was drawn at random from the same model LF used
above.  In this case, since we draw individual galaxies at random from the
parent LF, we also include random fluctuations in the LF itself.  For low to
moderate richness clusters, it is these fluctuations that dominate the scatter
in the fits to $M^*$ rather than the background correction method.  Although
the resultant values are therefore independent of redshift, we have adopted
$z=0.1$ as typical of our clusters, in order to include a suitable background
contribution.  This background is derived in the same manner as described
above.  Table 2 gives the limits outside of which 90\%, 31.5\%, 10\%, 5\% and
1\% of the data lie.  One of the most noticeable features of these simulations
is that the distribution of fitted values for $M^*$ is skew, with a strong tail
towards brighter magnitudes.  An example of this is shown in Figure 3, where we
present the distribution of fits to $M^*$ for two richnesses, $R_i=10$ and
$R_i=50$.  This trend is due to the restricted range of absolute magnitudes
that are actually used to fit against.  Since we set M$^*=-20.2$ in accordance
with the result we find in section 5.1, there are only two bins in the fitted
LF that actually sample the break.  At low richness values there are often
insufficient counts per bin to truly define this break, and hence, since there
is no constraint from the actual data, the fitted value of $M^*$ can take quite
large negative values without changing the $\chi^2$ of the fit significantly.
In practice, the limit for $M^*$ at the bright end is set by our constraint
that $M^*>-26$.  Clearly, for the richer sample shown in Figure 3 the trend
towards a strongly skewed distribution is much less evident as expected.  This
skewness also shows in the mean $M^*$ derived from the simulations, with this
value being the input value of $-20.2$ at large richness but diverging to
$-20.8$ at $R_i=10$.  This same behaviour is also demonstrated (in a slightly
different fashion) by the simulations of Colless (1989), where he shows that
the probability that a two-sample $\chi^2$ test can reject the possibility that
two clusters are drawn from the same underlying population is also skew with
respect to $M^*$.  We can further confirm the reason for this skewed
distribution, by looking at the distribution of $M^*$ when the absolute
magnitude range being fitted is $-23$ to $-18$.  The results for $R_i=25$ are
given in Table 2 and also shown in Figure 3.  It is easy to see the effect that
including the brighter galaxies has on fixing the location of the break in the
LF. The slight skewness left in the distribution now reflects the fact that the
total number of galaxies with $M<-21$ in these low richness systems is rather
few.  The derived error bounds including these bright galaxies are most similar
to those for a much higher value of $R_i$ ($\sim40$) when fitted over the
normal range in $M$.  The key result of these simulations therefore is that it
is difficult to rule against overly bright values of $M^*$ as being from the
same distribution as the composite.  However, those clusters for which $M^*$ is
rather fainter than the composite are strongly discriminated against.  We note
that the results of these simulations can also be applied to the data of
Colless (1989), and that the simulations of Dressler (1978) are not applicable
to the case of fitting over a restricted magnitude range (or rather that the
Dressler results would be closer to those achieved in practice but for a much
larger true richness).

It is worth also considering how sensitive these results are to our assumed
value of $\alpha$.  We tested this using $\alpha=1.2$ for $R_i\sim20$, typical
of our sample.  We found that the changes are small ($<0.1$ in the faint limit
on $M^*$, and $<0.2$ in the faint bright limit) and always act to push these
limits {\em brighter}.  Since the cluster LFs we can exclude are those which
have faint $M^*$, it is clear that adopting a lower value of $\alpha$ will only
make this difference larger.  The formal average of our composites using
all methods of calculating them give $\alpha=-1.22$ (Table 3).
This implies that the probability that any LF
agrees with the derived composite as given by the simulations is slightly
conservative.

Since we have carried out two parameter fits in section 5.1 and 5.3, we
also need to know the likely joint distribution of $\alpha$ and $M^*$ for the
case of rich systems alone (ie composite LFs).  We therefore carried out
simulations as described above but in this case applied a two parameter fit to
the LF.  In this instance the best way to present the data is graphically, and
the results of the simulations are shown in Figure 4.  The contour levels
plotted correspond to the same probabilities given for the one parameter case
in Table 2, except that we exclude the contour that contains 10\% of the data
in this instance.  The
simulations have been binned into 0.05 in $\alpha$ and 0.1 in $M^*$ for
presentation.  

It is clear from Figure 4 that $M^*$ and $\alpha$ are
highly correlated.  In what follows we will present formal error estimates
on each fit, derived by determining the value(s) at which 
$\chi^2 =\chi^2_{min}+1$.  
As will be seen, there is a considerable difference
between these formal errors and both the 
error distributions given in Table 2 for the one parameter
fits and the distribution shown in Figure 4 for the two parameter fits.  
This difference largely stems from
the correlation between $M^*$ and $\alpha$, since the errors in these
are also therefore clearly correlated.  These formal errors are included
for completeness only (since they do provide a measure of `goodness-of-fit').
We will only use the Monte-Carlo simulations and two-sample
$\chi^2$ tests when assessing the true deviations
that may be present between any given LF and a universal composite.

\section{Results}
\subsection{The Composite Luminosity Function}
The first LF we consider is the composite of all the clusters
which have $R_i>20$.  There are 23 clusters in this sample, and
the richness of the composite was found to be
$R_c \sim 800$ ($n^*n_{clus}\sim1600$).  This allows us to compare our 
results with previous samples which generally dealt with rich Abell clusters.

For this composite we have considered the effect of using the different
background correction methods outlined in Section 3.3, as well as
for the different assumed errors in that background subtraction.
The results of the fitting process for these
different background corrections are given in Table 3 and shown in Figure 5.

The first three rows were derived by fitting both $M^*$ and $\alpha$ to this
sample over the magnitude range ($-21\le m \le -18$) with an assumed
field-to-field error in the background counts of 50\%.  The first entry gives
the results for the local correction method, the second for the global
correction using all of the counts and the third the global correction but
using only the average of the annuli around the clusters.  Clearly the two
global background corrections we have used give essentially the same result.
Directly comparing the data derived from the local and global background
correction methods using a 2-sample $\chi^2$ test also shows that these
composites are in good agreement for the absolute magnitude range given above
($\gg$99\% probability they are same).  Moreover, since we have not
renormalised the LFs before comparing using this test, this also indicates that
differences that do exist between the background subtraction methods for
individual clusters (as will be shown in Section 5.2) tend to average out when
combined in the composite.

The last three rows in Table 3 give the equivalent results when the
field-to-field background error is assumed to be 25\% as noted in Section 3.3.
This change therefore gives the same results as before within the derived
errors on $M^*$ and $\alpha$ for the global correction method, and only very
slightly different in the formal sense for the local correction method.  This
shows that the dominant error in deriving estimates of $M^*$ or $\alpha$ is
actually the Poissonian contribution (cf the discussion after equation 4), and
that the variation due to assuming a different error on the background term
alone is small.  Indeed, the magnitude of these changes found here is
negligible compared to the expected measurement errors as shown in section 4.
Henceforth, we consider only the 50\% field-to-field variation, since it
adequately represents all of the clusters in the present sample.

Lastly, it is worth noting the effect that changing the magnitude range over
which the fits are made has on these conclusions.  Including the brighter
galaxies pushes $M^*$ brighter.  Including the fainter galaxies makes a slight
difference to the slope but little else.  It should be noted however that only
one cluster actually contributes for $M$ fainter than $-18$, whereas all 23
clusters contribute at $M$ brighter than $-21$.  The result derived when the
fit is over the range $-22<M<-17.5$, using the local background correction, is
also given in Table 3.  The low probability that the fit is good is due to the
fact that a Schechter function is now a poor fit to the data at both bright
{\em and } faint magnitudes.  We note that a similar (though larger) increase
in brightness of $M^*$ was found by Lugger when she considered mean luminosity
functions with the brightest cluster galaxies included.  However, we repeat our
caution of section 2.2 that data on such bright galaxies ($m<15$) cannot be
relied upon absolutely because of problems of saturation and image blending (cf
the discussion in Colless 1989).  It is also not clear therefore whether or not
the greater variance from the mean seen in the brightest bins, as compared to
those in the more restricted absolute magnitude range over which we normally
determine the LF, is actually real or merely an artifact of the COSMOS
measurements.

\subsection{Individual Cluster Luminosity Functions}
We also fitted each cluster LF separately to the Schechter function, but this
time holding $\alpha$ constant (since there are insufficient points in many of
the clusters to enable more than a one parameter fit to be made successfully).
We adopted $\alpha=-1.25$ as an
appropriate compromise, based both on our own results and those previously
published.   Table 4 shows the results of these tests for all 46
clusters that were found to have positive values of $R_i$ and Figure 6  shows
the actual data for those with $R_i\ge15$.  We quote results derived using both
the global and local background correction methods.  
 We also tested for differences arising from the choice of
background subtraction method by comparing the LF distributions directly.  We
used a two-sample $\chi^2$ test to check for differences between the two sets
of LFs, without rescaling the LFs for the possibility that the two background
subtraction methods may have resulted in a different overall normalisation.
We have allowed for the fact that the adopted magnitude completion limit
may be brighter than $M=-18$, so that we compare the LFs in the range $-21$ to
the brighter of $-18$ or the plate completion limit.  These limits are given in
Table 4.  The results of the two-sample $\chi^2$ test are given in column 10 of
Table 4.  As can be seen the agreement is good, and even for the worst case
(E748), we cannot rule out the possibility that the two samples represent the
same underlying distribution.

We also compared the single cluster LFs with a suitably normalised model of
the composite (assuming the average values of $M^*$ and $\alpha$ found above),
using a simple $\chi^2$ test (given in Table 4), 
and with the results of the Monte-Carlo
simulations from Table 2.  
It can be seen that there are several clusters where the LF
derived using global background subtraction is a poor match to the model, but
the LF derived using the local background is a much better fit.  We take this
as evidence for the fact that assuming an overall global background will not be
a good match to every cluster.  
Generally the two tests give similar results,
though we note that the two-sample  $\chi^2$ test is considerably more
conservative at ruling out individual LFs, as expected.  However, all 
those LFs for which the two-sample  $\chi^2$ test gives a probability 
below 20\%, are also found to be a poor match to the composite from the
simulations.  There are two exceptions to the corollary however:  both E438
and E462 have unexceptional probabilities from the two-sample  $\chi^2$ test,
but are excluded at the 5\% level from the simulations.

If we include only those clusters where the derived $M^*$ lies outside the 1\%
confidence level from the simulations (as taken from Table 2) for one or other
of the background subtraction methods, then only clusters E400, E429 and E524
show strong evidence for an LF that differs from the composite.  All three of
these clusters also disagree at the 5\% level when considering both background
subtraction methods.  At the 5\% level, there are several additional clusters
that disagree.  Leaving aside those where only one of the background
subtraction methods gives divergent results (E499, E726), we are left with
E438, E462 and E742.  All of these, except E400 and E429, have a value of $R_i
\le 15$, so we may reasonably exclude these as being different on the grounds
that accurate background subtraction is crucial for these clusters, and, as
found in Section 4, errors of 0.5 magnitudes in the fitted value of $M^*$ are
easily accounted for by an inappropriate background correction.  This leaves
only E400 and E429.  It can be seen from Table 4 that even the direct $\chi^2$
comparison of their LFs with the composite might lead to them being thought
discrepant.  The LFs of these clusters are apparently normal (see Figure 6).
However, before concluding that these clusters do show a significant
difference, we first consider the possibility that there may be some problem
with the data.  Examination of the data in Collins et al.\ (1995) shows that
all of these clusters have good reliable redshifts, with the positions of the
galaxies for which redshifts were obtained agreeing well with the cluster
centres given in Table 1.  Further, we note that for the cases of E400 and
E429, other clusters appear on the same photographic plate and do not have
discrepant $M^*$, hence the photometry of these systems should be good.
Therefore, we find that there are at least two clusters that do not have the
same LF as the derived composite.  This is in keeping with earlier studies by
Dressler (1978), Lugger (1986) and Colless (1989) who found that $\sim$10\% of
their clusters similarly did not agree with the mean Schechter LF they derived.

\subsection{Dependence of the LF on Cluster Properties}
The major advantage of our sample over previous studies is that it is large
enough that we can test the dependence of the LF on intrinsic cluster
properties.  In particular, we will consider the possible variations with
cluster richness (as defined by $R_i$), Bautz-Morgan class, which we have taken
from the Abell, Corwin \& Olowin catalogue, and velocity dispersion (taken from
Collins et al.\ 1995).

For each case we split our total sample into two, and used the local background
correction method only.  When comparing different richness samples we used the
cuts $10<R_i<20$ and $R_i>20$ (which is of course just the composite LF derived
previously).  For Bautz-Morgan class, we considered only those of type I and
III, and counted intermediate types (e.g.\ type I-II) as type II which we did
not use.  In this way we sought to maximise the differences between the
clusters to test for any effect on the LF.  We split the sample into two in
velocity dispersion using a cut at 700kms$^{-1}$.  Unfortunately, most of the
very high velocity dispersion clusters (those with $v>1000$kms$^{-1}$) are also
the most distant, making it difficult to constrain $\alpha$, and leaving few
bins in the data to compare the 2 sets using a $\chi^2$ test.  We therefore
were forced to use
a lower velocity cut-off for our high velocity dispersion sample than Colless
(1989) did, so do not sample the same part of parameter space as he did.  

Since we do not have well defined velocity dispersions for all clusters, and
those which are not in the Abell, Corwin \& Olowin catalogue do not have
Bautz-Morgan classes, the two subsamples do not comprise all of the observed
clusters.  The number of clusters comprising each subsample is given in Table
5.  In addition, for the velocity dispersion and Bautz-Morgan samples we have
included two richness cuts, at $R_i>20$ and $R_i>15$.  This provides some
indication of the stability of the result on the clusters making up the actual
subsamples.  We give both one and two parameter fits to the Schechter LF in
Table 5.  For the one parameter fits we again used $\alpha= -1.25$.  The
results from this analysis are also shown in Figure 7.  We have also compared
the LFs for these subsamples against the composite derived in Section 5.1, and
against each other, using a two-sample $\chi^2$ test, as well as
comparing the subsamples directly with
the results of our simulations (Figure 4).  The results of the two-sample
$\chi^2$ test are also given in Table 5.  Since $n^*$ can vary between these
different sets of clusters, we rescaled the binned data before the comparison,
so that the derived $R_c$ values from the composites agreed.

There are noticeable differences between the richness cuts for both the
Bautz-Morgan class III clusters and the high velocity dispersion clusters, with
the richer subsamples having a notably poorer fit to the standard Schechter LF,
and being discrepant compared to the composite.  However, only one cluster
contributes to the faintest bin in the richer subsample, and only one other is
added to that when the poorer limit is considered.  If we exclude this bin,
then very similar fits are found for both richness limits and the discrepancy
is removed.  It is clear from this that variations in the LF due to the last
bin alone are not significant.  We do not consider these two subsamples
further.

Once these discrepant sets are removed, it is clear that neither Bautz-Morgan
class shows any evidence for deviation from a universal LF, nor do they differ
from each other, according to any of the tests applied.  The same is true for
the split according to richness alone, where the fits to both samples are very
similar within the derived errors.  The actual `goodness' of the fit for the
lowest richness sample is poor, but this most likely reflects difficulties in
background correction at these low richness values as noted previously.

For the $R_i>15$ velocity dispersion samples, there is weak evidence that the
two samples differ from each other (significance $\sim54$\%).  When compared
with the composite, neither sample shows any evidence for difference when using
the two-sample $\chi^2$ test.  However, when compared with the Monte-Carlo
simulations, we find that the higher velocity dispersion clusters are clearly
unlikely to have been drawn from the same basic distribution as the composite
(at much better than the 1\% level).  We therefore find that there may be
evidence for some dependence of the LF on velocity dispersion.

It is worth considering the effect of including the brighter galaxies as well
in this analysis.  As is readily evident from Figure 7, there is little
difference between the rich and poor samples across the entire range of $M$,
and the differences for the high and low velocity dispersion samples is largely
restricted to the fainter galaxies.  Considering the $R_i>15$ Bautz-Morgan
samples however, there are clearly fewer bright galaxies in the Bautz-Morgan
class I sample than in the Bautz-Morgan class III sample.  This is not a
normalisation effect since the two samples have essentially the same LFs at
fainter magnitudes.  Again however, we caution against overinterpretation of
this difference.

Lastly, we note that all of the velocity dispersion subsets, and the low
richness subset can formally be excluded as being drawn from a model of the
composite LF with $\alpha$ fixed at $-1.25$ on the basis of the one parameter
fits.  Small deviations in both $M^*$ and $\alpha$ mean that the one parameter
fits are distinct from the composite derived in section 5.1.  It is clearly
dangerous to compare such subsamples (with their relatively small intrinsic
errors) using such methods.

\section{Discussion}
\subsection{Comparison with Previous Results}
First, we compare our derived estimates for the composite LF,
for all clusters with $R_i>20$, with those
derived previously.  Colless (1989) gives the best fits from previous data in
his Table 2,  scaled into the colour and value of $H_0$ that we are using.  
Excluding Colless' own data this shows that the quoted values of $M^*$ and
$\alpha$ are $-19.9$ and $-1.24$ respectively, with typical errors in $M^*$ of
0.5,  when the brightest cluster galaxies are excluded from the fit. Colless
himself finds $M^*=-20.04$ and $\alpha=-1.21$ for a two parameter fit similar
to ours. These  values are in good agreement within the errors 
with those derived by us.  
It is clear that the luminosity functions of moderately rich clusters,
when considered across this relatively narrow range in absolute magnitude,
are indeed very similar on average.

In addition we can compare our one parameter fits for 3 clusters with those
given by Colless (1989).  EDCC clusters 394, 400 and 124 correspond to his
clusters C02, C03 and C52.  From Table 4 we can see that when a comparison is
made of similar fitting procedures (ie, $\chi^2$ fitting and using a global
background correction), the agreement is excellent.  Colless quotes values for
$M^*$ of $-19.76\pm0.21$, $-19.86\pm0.23$ and $-20.05\pm0.33$ for C02, C03 and
C52 respectively.  Although the original photographic plates used in both
studies are the same, a different measuring machine was used (the APM), with
consequent differences in the reduction from measuring machine magnitudes to
final derived astronomical magnitudes.  Therefore, this agreement for the
specific clusters that are common to the two studies is highly encouraging.  It
is worth making some specific comments on these clusters however.  E394 was
found by Colless to be discrepant with both the fit by a Schechter LF and with
a comparison with his composite LF.  We note that Table 4 shows that whilst the
global background correction gives a moderate probability for both the fit and
the comparison with the composite, the local background correction method gives
better agreement.  This difference shows in the direct comparison of the
magnitude distributions for the two background correction methods (Table 4).
This is clearly an example where global background correction gives rise to
these deviations.  We note that E394 is one of the poorest clusters in both our
sample and that of Colless.  Colless found that E124 was marginally discrepant
when compared against the composite LF.  We find no evidence for any
difference.  Lastly, E400, which we found in Section 5.2 to be potentially
significantly different from the composite, Colless finds to be a reasonable
 match.  However, since this cluster gives a difference for both background
subtraction methods, we find our result to be reliable.  We also note
that the smaller errors derived for by us for both E400 and E124 allow
us to make more definitive statements than Colless can for these clusters.

Finally we note that unlike Colless, we find that all the subsets of the data
we took are well fit by a one parameter Schechter LF, though they are not
necessarily in agreement with the fit to the composite.

\subsection{Implications for cluster formation models}
There have been few detailed N-body simulations of the formation
of clusters that have suitable dynamic range to map out the whole of the
expected LF as a function of the cosmological model assumed.  
For example, van Kampen (1995) makes simple predictions 
for the LF based on extensive simulations, but is led to the conclusion
that his modelling cannot correctly account for all of the merging that
takes place within the cluster.  His fit to the bright end of the luminosity
function (which we might expect to be easier to model) is similar to the
one observed.

There have been studies that lead to simpler predictions as to the expected
form of the LF according to the dominant processes that occur during cluster
formation.  The mode of cannabalism for the growth of cD galaxies as suggested
by Hausman \& Ostriker (1978) predicts that there should be differences between
the luminosity function when broken down by Bautz-Morgan class.  We see no real
evidence for such a difference when we consider the range $-18>M>-21$.
There is an indication for such a trend in the galaxies brighter than
$M=-21$, in the sense that the Bautz-Morgan class I sample has fewer bright
galaxies.  However, better photometry of the bright galaxies would be required
to confirm this result.  Merritt (1983,1984) argued that
most of the properties of clusters are essentially `frozen in' when they form
and that merging and tidal stripping processes are ineffective in rich cluster
cores.  This leads to the prediction that the LF should appear to be much the
same from cluster to cluster.  However, HST imaging data of moderate redshift
clusters (e.g.\ Dressler et al.\ 1994) tends to rule against such a model, since
it provides convincing evidence that mergers are important at least in the
early stages of cluster evolution.  One possibility is that the bulk of the
change in the LF between clusters occurs only amongst the very faintest or
brightest galaxies, the regime to which our data is insensitive.  For example,
it is known that some clusters show an upturn in the LF at faint magnitudes
(e.g.\ Driver et al.\ 1994, Bernstein et al.\ 1995), and 
there has always been marginal
evidence for greater deviations in the LF when the brightest cluster galaxies
are included (e.g.\ Lugger 1986 and the comments in Section 5.1), though
we again caution against over-interpretation of results on the brightest
cluster members.  However, our data clearly shows that in the region of the
`break' in the LF, the cluster-to-cluster variance is extremely small.

\section{Conclusions}
We have considered the cluster galaxy luminosity function as derived from data
in the Edinburgh-Durham Southern Galaxy Catalogue and the Edinburgh-Milano
Redshift Survey.  Our dataset is the largest multi-redshift cluster sample to
be considered to date for this purpose.  From this we find:
\begin{enumerate}
\item
Our data shows that there is strong evidence that {\em
most} clusters have very similar LFs when both the brightest and faintest
galaxies are discarded.  The average of the composite LFs derived using the 3
background correction methods in Section 5.1 give $M^*=-20.16\pm0.02$ and
$\alpha=-1.22\pm0.04$, when the LF is determined in the range $-18>M>-21$.
\item We find a very good agreement between our results and those 
 derived by Colless (1989) from similar input data, and with the mean of
the data presented by Dressler (1978) and Lugger (1986).  
\item
 We have tested the
stability of the derived Schechter luminosity function parameters to the type
of background correction made.  We find excellent agreement between a method
based on a global background correction and that based on a strictly local
correction.  However, the largest systematic error remaining in our derivation
of the best fit parameters for the composite LFs is the background
correction, with an uncertainty of up to 0.05 in $M^*$ and 0.05 in $\alpha$
possible.  
\item
We  also tested the universality of the luminosity function, by comparing
individual cluster LFs with the composite function derived in section 5.1, and
by testing each cluster LF separately against a Schechter function (section
5.2).  From this we find that at most $\sim$10\% of the clusters may have LFs
that are significantly different from the composite.  
\item We also broke our sample into sub-samples defined by richness,
Bautz-Morgan class and velocity dispersion (section 5.3).  From these tests we
found weak evidence that the high and low velocity dispersion samples have
different composite LFs, and that the higher velocity dispersion clusters
may have a different LF to the global composite.  
This agrees with the marginal detection of a
difference between similar samples by Colless (1989).  We also found evidence
for differences between the Bautz-Morgan class samples, but only for the
brightest galaxies ($M<-21$) that we otherwise excluded for reasons of
photometric reliability (Section 3.4).  The trend evident is in line with the
predictions of the cannabalism model of Hausman \& Ostriker (1978), but we
caution against over-interpretation of this result without more reliable
photometry for the bright galaxies.  There is no evidence for any convincing
difference between the other subsets of the data.  Our results, and our
simulations, do however show that it is important to consider more than one
method of testing the difference between subsets of the data.  On the basis of
a one parameter fit alone we would have concluded that there were significant
differences between the low and high velocity dispersion samples and the
composite, and between the low and high richness clusters.  The other tests
show no evidence for any difference however, with the exception of the high
velocity dispersion clusters where the two parameter fits, and Monte-Carlo
simulations show weak evidence for a difference from the composite.
\end{enumerate}

It is likely that to make any further progress on the nature of the cluster
galaxy luminosity function we will need to overcome the  limitations of
data such as ours.  First, there is clearly a need to derive cluster membership
free of any correction for backgrounds, by obtaining sufficient redshifts that
we can actually map the galaxy distribution in three dimensions around the
cluster.  Secondly, better photometry is required to tie down the bright and
faint ends of the LF. Given
these it should be possible to not only discriminate between differences in the
data broken down by richness, velocity dispersion or whatever else is desired,
but also to test for the differences that our data suggest
may exist in the LF when comparing
the very brightest and very faintest galaxies.

\section*{Acknowledgements}
SLL would like to acknowledge the support of an SERC Postdoctoral Fellowship
during the early part of his work on the EDCC.  CAC acknowledges the
support of PPARC through an Advanced Fellowship.  We would like to thank
the referee, Steve Maddox, for his comments.

\parindent=0pt

\section*{References}
\mnref{Abell, G.O., 1975, in Sandage, A., Sandage, M., Kristian, J., eds,
	Stars and Stellar Systems IX: Galaxies and the Universe, Univeristy
	of Chicago, Chicago, p. 601}
\mnref{Abell, G.O., Corwin, H.G., Olowin, R.P., 1989, \apjs, 70, 1}
\mnref{Bernstein, G.M., Nichol, R.C., Tyson, J.A., Ulmer, M.P., 
	Wittman, D., 1995, \aj, 110, 1507}
\mnref{Colless, M., 1989, \mn, 237, 799}
\mnref{Collins, C.A., Nichol, R.C., Lumsden, S.L., 1992, 
      \mn,  254, 295.}
\mnref{Collins, C.A., Guzzo, L., Nichol, R.C. \& Lumsden, S.L., 1995, 
      \mn, 274, 1071.}
\mnref{Dressler, A., 1978, \apj, 223, 765}
\mnref{Dressler, A., Oemler, A., Sparks, W.B., Lucas, R.A., 1994, 435, L23} 
\mnref{Driver, S.P., Phillipps, S.P., Davies, J.I>, Morgan, I., Disney, M.J.,
	1994, \mn, 268, 393}
\mnref{Ellis, R.S., 1983, in Jones, B.J.T., Jones, J.E., eds, The Origin and
	Evolution of Galaxies, Reidel, Dordrecht, p. 225}
\mnref{Geller, M.J., Beers, T.C., PASP, 94, 421}
\mnref{Guzzo, L., Collins, C.A., Nichol, R.C., Lumsden, S.L., 1992, \apj, 393,
	 L5}
\mnref{Hausman, M.A., Ostriker, J.P., 1978, \apj, 224, 320}
\mnref{Heydon-Dumbleton, N.H., 1989, PhD Thesis, University of Edinburgh}
\mnref{Heydon-Dumbleton, N.H., Collins, C.A., MacGillivray, H.T., 1989, \mn,
	238, 379}
\mnref{Lugger, P., 1986, \apj, 303, 535}
\mnref{Lumsden, S.L., Nichol, R.C., Collins, C.A., Guzzo, L., 1992, 
      \mn,  258, 1.}
\mnref{Maddox, S.J., Sutherland, W.J., Efstathiou, G.P., Loveday, J., 
  Peterson, B.A., 1990, \mn, 247, 1}
\mnref{Malumuth, E., Richstone, D.O., 1984, \apj, 276, 413}
\mnref{Merritt, D., 1983, \apj, 264, 24}
\mnref{Merritt, D., 1984, \apj, 276, 26}
\mnref{Merritt, D., 1985, \apj, 289, 19}
\mnref{Metcalfe, N., Shanks, T., Fong, R., Jones, L.R., 1991, \mn,
	249, 498}
\mnref{Nichol, R.C., Collins, C.A., Guzzo, L., Lumsden, S.L., 1992, 
      \mn,  255, 21P.}
\mnref{Nichol, R.C., Collins, C.A., 1993, \mn, 265, 867}
\mnref{Richstone, D.O., 1975, \apj, 200, 535}
\mnref{Richstone, D.O., 1976, \apj, 204, 642}
\mnref{Schechter, P.L., 1976, \apj, 203, 297}
\mnref{van Kampen, E., 1995, \mn, 273, 295}

\onecolumn

\begin{table}
\begin{tabular}{r rrr rrr r l rrc rrr}
EDCC & \multicolumn{3}{c}{RA} & \multicolumn{3}{c}{Dec} & Field & Redshift 
& N$_{tot}$ & N$_{clus}$ & BM class & \multicolumn{3}{c}{Vel. Disp.
(km/s)}  \\
 42 & 21 & 46 & 21.9 & -30 & 56 & 37.8 & 466 & 0.11949 & 8 & 4 & 2 \\
 51 & 21 & 49 & 22.2 & -29 & 8 & 1.7 & 466 & 0.0927 & 6 & 5 & 0 \\
 61 & 21 & 53 & 59.4 & -30 & 19 & 37.8 & 466 & 0.09257 & 4 & 4 & 0 \\
 124 & 22 & 14 & 43.9 & -35 & 57 & 33.1 & 405 & 0.14661 & 10 & 9 & 2 & 
\phantom{disper}& 1011 \\
 131 & 22 & 16 & 39.2 & -34 & 56 & 27 & 405 & 0.15711 & 4 & 4 & 3 \\
 145 & 22 & 24 & 56.7 & -30 & 51 & 11 & 467 & 0.05697 & 12 & 11 & 2 && 1105 \\
 261 & 23 & 9 & 9.4 & -29 & 19 & 41.1 & 469 & 0.11709 & 8 & 6 & 1 && 820 \\
 311 & 23 & 28 & 36 & -36 & 47 & 41.4 & 408 & 0.09544 & 7 & 5 & 1 \\
 348 & 23 & 44 & 33.7 & -28 & 31 & 40.6 & 471 & 0.0292 & 33 & 32 & 3 && 830 \\
 366 & 23 & 52 & 19.6 & -27 & 56 & 40 & 471 & 0.07278 & 8 & 6 & 3 && 464 \\
 392 & 0 & 0 & 13.9 & -34 & 56 & 38.4 & 349 & 0.11272 & 8 & 7 & 3 && 525 \\
 394 & 0 & 0 & 32.1 & -36 & 12 & 58.9 & 349 & 0.04902 & 6 & 6 & 1 && 598 \\
 400 & 0 & 3 & 39.1 & -34 & 58 & 49.1 & 349 & 0.11307 & 10 & 7 & 2 && 1222 \\
 408 & 0 & 7 & 27.8 & -35 & 56 & 8.6 & 349 & 0.11936 & 14 & 13 & 2 && 871 \\
 421 & 0 & 13 & 35.9 & -35 & 13 & 53.1 & 350 & 0.14618 & 7 & 5 & 3 \\
 429 & 0 & 15 & 23.2 & -35 & 25 & 2.5 & 350 & 0.09693 & 19 & 17 & 3 && 790 \\
 437 & 0 & 18 & 1.2 & -25 & 54 & 26.3 & 473 & 0.1432 & 6 & 5 & 1 \\
 438 & 0 & 20 & 23.5 & -38 & 24 & 12.6 & 294 & 0.11919 & 4 & 4 & 2 \\
 450 & 0 & 27 & 23.4 & -29 & 45 & 1.3 & 410 & 0.0988 & 5 & 4 & 3 \\
 460 & 0 & 34 & 47.1 & -28 & 44 & 47.4 & 411 & 0.1122 & 31 & 21 & 1 && 700 \\
 462 & 0 & 35 & 14.4 & -39 & 23 & 42.4 & 294 & 0.06316 & 12 & 11 & 1 && 569 \\
 470 & 0 & 37 & 26.4 & -26 & 26 & 25.1 & 474 & 0.10975 & 10 & 7 & 0 && 415 \\
 471 & 0 & 37 & 43.3 & -24 & 56 & 48.3 & 474 & 0.11175 & 7 & 4 & 3 \\
 473 & 0 & 40 & 3.7 & -28 & 50 & 23.2 & 411 & 0.10799 & 14 & 13 & 1 && 695 \\
 474 & 0 & 40 & 44.7 & -26 & 19 & 53.9 & 474 & 0.11256 & 15 & 8 & 3 && 354 \\
 482 & 0 & 46 & 50.3 & -29 & 47 & 22 & 411 & 0.10783 & 30 & 21 & 1 && 675 \\
 485 & 0 & 48 & 56.3 & -28 & 46 & 50.4 & 411 & 0.11251 & 6 & 6 & 2 && 443 \\
 495 & 0 & 53 & 28.6 & -26 & 36 & 9.4 & 474 & 0.11412 & 17 & 14 & 3 && 403 \\
 499 & 0 & 53 & 51.4 & -38 & 10 & 2.1 & 295 & 0.11697 & 4 & 4 & 3 \\
 519 & 1 & 2 & 7.7 & -40 & 6 & 22.4 & 295 & 0.1073 & 13 & 9 & 1 && 371 \\
 524 & 1 & 5 & 39.9 & -37 & 1 & 20.6 & 352 & 0.11751 & 9 & 9 & 1 && 975 \\
 553 & 1 & 23 & 9 & -39 & 41 & 37.1 & 296 & 0.08791 & 14 & 12 & 1 && 242 \\
 557 & 1 & 23 & 46.9 & -38 & 14 & 34.8 & 296 & 0.07969 & 8 & 6 & 1 && 413 \\
 575 & 1 & 31 & 52.4 & -27 & 47 & 19.3 & 476 & 0.12545 & 7 & 5 & 1 \\
 606 & 1 & 58 & 27.2 & -33 & 11 & 15.2 & 354 & 0.10045 & 12 & 11 & 3 && 1152 \\
 653 & 2 & 27 & 16.9 & -33 & 41 & 37 & 355 & 0.07924 & 6 & 6 & 0 && 440 \\
 658 & 2 & 28 & 34.9 & -33 & 17 & 55.5 & 355 & 0.07636 & 26 & 22 & 3 && 977 \\
 683 & 2 & 42 & 26.6 & -26 & 31 & 10.7 & 479 & 0.13364 & 4 & 4 & 3 \\
 699 & 2 & 49 & 17.9 & -25 & 9 & 1.2 & 480 & 0.11113 & 7 & 6 & 2 && 891 \\
 712 & 2 & 54 & 19.6 & -24 & 55 & 51.1 & 480 & 0.11093 & 19 & 13 & 1 && 519 \\
 722 & 3 & 1 & 3.9 & -37 & 7 & 47.2 & 357 & 0.06664 & 4 & 4 & 1 \\
 726 & 3 & 4 & 43 & -39 & 1 & 47.3 & 300 & 0.08737 & 4 & 4 & 0 \\
 728 & 3 & 6 & 13.3 & -36 & 53 & 32.2 & 357 & 0.06748 & 9 & 8 & 1 && 323 \\
 735 & 3 & 9 & 23.7 & -27 & 5 & 34.3 & 481 & 0.06826 & 9 & 7 & 1 && 359 \\
 742 & 3 & 11 & 52.2 & -38 & 30 & 34.7 & 300 & 0.08384 & 6 & 6 & 1 && 709 \\
 748 & 3 & 13 & 9.5 & -29 & 24 & 14.2 & 417 & 0.06709 & 6 & 5 & 2 \\
\end{tabular}
\caption{The full cluster sample.  The EDCC number refers to the identification
in Lumsden et al.\ (1992).  The field is the original Schmidt J survey field
number.  N$_{tot}$ is the actual number of redshifts obtained for that cluster,
and N$_{clus}$ is the number of those redshifts adjudged to lie in the cluster.
BM class is the Bautz-Morgan class, where 1 represents class I, 3 represents
class III, 2 represents any other defined class and 0 implies that the cluster
concerned is not in the Abell et al.\ (1989) catalogue and therefore has not
been classified.  The velocity dispersion is taken directly from Collins et
al.\ (1995).  }
\end{table}

										\begin{table}
\begin{tabular}{lcccccccccc}
$R_i$ & \multicolumn{2}{c}{90\%} & \multicolumn{2}{c}{31.5\%}
	 & \multicolumn{2}{c}{10\%} & \multicolumn{2}{c}{5\%} 
	& \multicolumn{2}{c}{1\%} \\
200 & $-20.21$ & $-20.18$ & $-20.32$ & $-20.09$ & $-20.41$ & $-20.03$ &
	$-20.45$ & $-20.01$ & $-20.56$ & $-19.95$ \\
50 & $-20.24$ & $-20.18$ & $-20.49$ & $-19.98$ & $-20.74$ & $-19.87$ & $-20.89$
	& $-19.82$ & $-21.36$ & $-19.74$ \\
40 & $-20.24$ & $-20.16$ & $-20.57$ & $-19.95$ & $-20.90$ & $-19.83$ & $-21.17$
        & $-19.76$ & $-22.00$ & $-19.65$ \\
35 & $-20.25$ & $-20.17$ & $-20.60$ & $-19.95$ & $-20.91$ & $-19.83$ & $-21.22$
        & $-19.75$ & $-22.31$ & $-19.62$ \\
30 & $-20.28$ & $-20.19$ & $-20.68$ & $-19.94$ & $-21.19$ & $-19.78$ & $-21.70$
        & $-19.74$ & $*$ & $-19.59$ \\
25 & $-20.29$ & $-20.17$ & $-20.79$ & $-19.87$ & $-21.55$ & $-19.71$ & $-21.99$
        & $-19.63$ & $*$ & $-19.50$ \\
20 & $-20.32$ & $-20.19$ & $-21.06$ & $-19.88$ & $-22.71$ & $-19.73$ & $*$
        & $-19.62$ & $*$ & $-19.51$ \\
15 & $-20.35$ & $-20.18$ & $-21.32$ & $-19.84$ & $-25.94$ & $-19.62$ & $*$
        & $-19.55$ & $*$ & $-19.14$ \\
10 & $-20.35$ & $-20.18$ & $-21.39$ & $-19.75$ & $*$ & $-19.45$ & $*$
        & $-19.34$ & $*$ & $-19.04$ \\
25 & $-20.23$ & $-20.17$ & $-20.52$ & $-20.00$ & $-20.88$ & $-19.88$ & $-21.11$
        & $-19.83$ & $-21.73$ & $-19.70$ \\

\end{tabular}
\caption{The measured distribution of the fitted values of $M^*$ as a function
of richness from Monte-Carlo simulations.  The percentage values listed
refer to the values of $M$ that exclude that fraction of the actual measured
fits.  Since the distribution is skew, two limits are given for each
percentage.  A $*$ implies that no acceptable fits could be found ($M^*<-26$)
for that fraction of the data.
The last row gives the results for a simulation in which data
was fitted over the range $-23$ to $-18$, showing the effect that the
truncation at the bright end of the LF has on the fitting process.}
\end{table}

\begin{table}									
\begin{tabular}{lllrrrr}	
Sample & \multicolumn{1}{c}{$M^*$} & \multicolumn{1}{c}{$\alpha$}
 & p(fit) & $n_{clus}$ & $n_{clus}n^*$ & $R_C$\\[0.2in]
50\% Background error\\[0.1in]

Local background correction 
&  $-20.14\pm0.01$ & $-1.18\pm0.01$ & 92 & 23 & 1660 & 770\\
Global background correction
&  $-20.19\pm0.01$ & $-1.25\pm0.01$ & 99 & 23 & 1620 & 810\\
Global correction using the annuli
       & $-20.15\pm0.01$  & $-1.23\pm0.01$ & 97 & 23 & 1650 & 790\\[0.1in]

All galaxies in the range $-22<M<-17.5$, 
&  $-20.66\pm0.07$  & $-1.27\pm0.03$ & 2 & 23 & 970 & 770\\
local correction\\[0.2in]

25\% Background error\\[0.1in]

Local background correction 
&  $-20.09\pm0.02$ & $-1.16\pm0.01$ & 86& 23 & 1750 & 770\\
Global background correction
&  $-20.19\pm0.01$ & $-1.25\pm0.01$ & 97 & 23& 1620& 810\\
Global correction using the annuli
       & $-20.14\pm0.01$  & $-1.22\pm0.01$ & 94& 23 & 1670 & 790\\
\end{tabular}
\caption{Fits to the various composite LFs described in the text.
The  columns are:
 the description of the composite; the
derived values of $M^*$and $\alpha$ (the error corresponds to the values
of $M^*$ and $\alpha$ where $\chi^2 = \chi^2_{min}+1$);  
the formal likelihood that the
fit is good; the number of clusters used in constructing the composite;
the derived value of $n^*$; and the measured richness of the composite.}
\end{table}

\begin{table}
\begin{tabular}{rccrcccrcrc}
EDCC  & $M^*$ (global) & p(fit) & $R$ & p(CLF) & $M^*$ (local) & p(fit) & 
      $R$ & p(CLF) & p(same) & $M$(faint) \\
 42  & $-20.68\pm0.18$ & 70 & 61 & 47 & $-20.53\pm0.24$ & 85 & 49 & 85 & 99 & $-18.74 $ \\
 51  & $-20.60\pm0.35$ & 96 & 11 & 44 & $-20.71\pm0.69$ & 76 &  7 & 36 & 86 & $-18.07 $ \\
 61  & $-19.85\pm0.16$ & 24 & 33 & 21 & $-20.55\pm0.36$ & 89 & 25 & 95 & 98 & $-18.04 $ \\
 124 & $-19.94\pm0.12$ & 52 & 53 & 57 & $-19.95\pm0.14$ & 62 & 48 & 67 &100 & $-18.90 $ \\
 131 & $-20.13\pm0.20$ & 94 & 37 & 99 & $-20.23\pm0.23$ & 95 & 36 &100 & 99 & $-19.09 $ \\
 145 & $-20.42\pm0.24$ & 57 & 18 & 67 & $-20.42\pm0.24$ & 62 & 19 & 77 & 93 & $-18.00 $ \\
 261 & $-20.13\pm0.22$ & 23 & 25 & 34 & $-20.17\pm0.21$ & 24 & 28 & 39 & 98 & $-18.38 $ \\
 311 & $-20.67\pm0.28$ & 85 & 28 & 70 & $-20.70\pm0.31$ & 84 & 27 & 71 &100 & $-18.00 $ \\
 348 & $-19.35\pm0.15$ & 73 & 10 &  9 & $-19.25\pm0.19$ & 74 &  8 & 15 & 99 & $-18.00 $ \\
 366 & $-20.45\pm0.36$ & 94 & 12 & 90 & $-19.92\pm0.45$ & 66 &  8 & 72 & 94 & $-18.00 $ \\
 392 & $-20.70\pm0.26$ & 95 & 29 & 90 & $-20.71\pm0.43$ & 99 & 22 & 98 &100 & $-18.19 $ \\
 394 & $-19.74\pm0.23$ & 24 &  9 & 20 & $-20.18\pm0.41$ & 31 &  8 & 54 & 52 & $-18.00 $ \\
 400 & $-19.64\pm0.11$ & 62 & 39 &  8 & $-19.58\pm0.14$ & 70 & 33 & 19 &100 & $-18.20 $ \\
 408 & $-20.40\pm0.28$ & 53 & 17 & 34 & $-21.42\pm0.66$ & 60 & 12 & 39 & 63 & $-18.35 $ \\
 421 & $-20.07\pm0.18$ & 64 & 36 & 80 & $-20.06\pm0.22$ & 62 & 31 & 81 & 98 & $-18.91 $ \\
 429 & $-19.47\pm0.13$ & 71 & 22 &  3 & $-19.55\pm0.13$ & 73 & 22 & 13 &100 & $-18.00 $ \\
 437 & $-20.26\pm0.16$ & 97 & 53 & 99 & $-20.30\pm0.15$ & 99 & 56 & 99 &100 & $-18.93 $ \\
 438 & $-19.49\pm0.24$ & 91 & 15 & 54 & $-19.47\pm0.24$ & 94 & 15 & 50 & 99 & $-18.57 $ \\
 450 & $-21.47\pm0.53$ & 50 & 18 & 45 & $-21.14\pm0.51$ & 72 & 18 & 78 & 83 & $-18.00 $ \\
 460 & $-20.56\pm0.26$ & 27 & 35 & 19 & $-20.33\pm0.27$ & 30 & 32 & 29 & 96 & $-18.73 $ \\
 462 & $-19.22\pm0.27$ & 69 &  9 & 51 & $-19.18\pm0.20$ & 49 & 10 & 23 &100 & $-18.00 $ \\
 470 & $-20.93\pm0.56$ & 80 & 13 & 74 & $-19.68\pm0.40$ & 89 & 10 & 85 & 68 & $-18.35 $ \\
 471 & $-20.83\pm0.34$ & 98 & 26 & 83 & $-20.50\pm0.32$ & 97 & 24 & 96 & 98 & $-18.39 $ \\
 473 & $-20.52\pm0.17$ & 86 & 45 & 82 & $-20.32\pm0.17$ & 91 & 42 & 96 & 98 & $-18.63 $ \\
 474 & $-20.82\pm0.37$ & 22 & 23 & 30 & $-20.43\pm0.30$ & 16 & 23 & 26 & 99 & $-18.40 $ \\
 482 & $-20.36\pm0.25$ & 66 & 24 & 73 & $-20.11\pm0.27$ & 77 & 20 & 85 & 98 & $-18.63 $ \\
 485 & $-20.17\pm0.27$ & 54 & 19 & 68 & $-20.10\pm0.28$ & 62 & 18 & 78 &100 & $-18.73 $ \\
 495 & $-20.04\pm0.15$ & 95 & 40 & 97 & $-20.11\pm0.13$ & 97 & 45 & 99 &100 & $-18.47 $ \\
 499 & $-19.57\pm0.27$ & 90 & 14 & 81 & $-19.68\pm0.25$ & 94 & 16 & 85 & 99 & $-18.45 $ \\
 519 & $-20.46\pm0.26$ & 93 & 22 & 96 & $-20.44\pm0.25$ & 95 & 23 & 96 &100 & $-18.00 $ \\
 524 & $-18.96\pm0.17$ & 78 & 13 &  9 & $-19.09\pm0.16$ & 77 & 14 & 12 & 94 & $-18.35 $ \\
 553 & $-19.92\pm0.25$ & 93 & 15 & 93 & $-19.76\pm0.27$ & 99 & 13 & 96 &100 & $-18.00 $ \\
 557 & $-19.57\pm0.18$ & 77 & 15 & 30 & $-19.47\pm0.19$ & 69 & 13 & 18 & 99 & $-18.00 $ \\
 575 & $-20.35\pm0.23$ & 43 & 20 & 14 & $-20.42\pm0.20$ & 32 & 23 &  7 &100 & $-18.51 $ \\
 606 & $-21.26\pm0.51$ & 41 & 12 & 30 & $-21.13\pm0.41$ & 53 & 15 & 35 & 90 & $-18.00 $ \\
 653 & $-19.84\pm0.33$ &100 & 10 & 96 & $-19.62\pm0.26$ & 97 & 10 & 77 & 97 & $-18.00 $ \\
 658 & $-19.27\pm0.36$ & 77 &  5 & 45 & $-19.16\pm0.24$ & 93 &  6 & 28 & 99 & $-18.00 $ \\
 683 & $-20.79\pm0.25$ & 51 & 42 & 46 & $-20.77\pm0.23$ & 58 & 43 & 52 &100 & $-18.70 $ \\
 699 & $-19.89\pm0.15$ & 82 & 35 & 83 & $-19.90\pm0.14$ & 82 & 37 & 81 &100 & $-18.21 $ \\
 712 & $-23.95\pm1.00$ & 63 & 10 & 50 & $-21.60\pm0.68$ & 52 & 12 & 46 & 99 & $-18.21 $ \\
 722 & $-20.32\pm0.22$ & 62 & 28 & 67 & $-20.48\pm0.29$ & 70 & 25 & 68 & 98 & $-18.00 $ \\
 726 & $-19.77\pm0.21$ & 70 & 15 & 52 & $-19.30\pm0.30$ & 82 & 10 & 43 & 88 & $-18.00 $ \\
 728 & $-20.22\pm0.23$ & 51 & 19 & 62 & $-19.98\pm0.31$ & 77 & 14 & 88 & 99 & $-18.00 $ \\
 735 & $-19.42\pm0.18$ & 79 & 12 & 17 & $-19.42\pm0.16$ & 88 & 13 &  8 & 99 & $-18.00 $ \\
 742 & $-19.57\pm0.16$ & 30 & 17 & 12 & $-19.40\pm0.25$ & 33 & 11 & 12 & 82 & $-18.00 $ \\
 748 & $-19.76\pm0.30$ & 61 & 10 & 68 & $-19.92\pm0.21$ & 42 & 12 & 28 & 10 & $-18.00 $ \\
\end{tabular}									
\caption{Best fit values of $M^*$ for a one parameter Schechter function fit to
the individual cluster LFs.  The error quoted corresponds to the mean of the
deviations from $M^*$ where $\chi^2 = \chi^2_{min}+1$.  The probability quoted
is the level at which this fit can be accepted.  Both global and local
background subtraction methods are given.  For both of these the probability
from a 2-sample $\chi^2$ test that the actual data differs from the composite
LF is also given as p(CLF). Similarly, The probability that the local and
global background subtracted data are the same is given as p(same).  Lastly,
the faint magnitude limit used in these tests is given. The bright limit is
always $-21$. }
\end{table}									

\begin{table}									
\begin{tabular}{lllrrrrr}							
Sample & \multicolumn{1}{c}{$M^*$} & \multicolumn{1}{c}{$\alpha$}
 & p(fit) & $n_{clus}$ & $n_{clus}n^*$ & $R_c$ & 		
 p(CLF)\\									
BM class I ($R_i>20$)
&  $-20.44\pm0.14$ & $-1.28\pm0.04$ & 69 & 8 & 400 & 260 & $>99$\\
           &  $-20.37\pm0.08$ & $-1.25$ & 83 & & 430 & \\
BM class I ($R_i>15$)
&  $-20.35\pm0.14$ & $-1.24\pm0.05$ & 63 & 9 & 480 & 280 & $>99$\\
           &  $-20.37\pm0.08$ & $-1.25$ & 78 & & 470 & \\

BM class III ($R_i>20$)
       & $-22.60\pm5.80$  & $-2.00\pm0.03$ & 79 & 8 & 21 & 250 & 16\\
            & $-19.82\pm0.10$ &$-1.25$ & 12 & & 860 & \\
BM class III ($R_i>15$)
      & $-20.56\pm0.18$  & $-1.49\pm0.04$ & 74 & 10 & 360 & 280 & $>99$\\
            & $-20.09\pm0.07$ &$-1.25$ & 77 & & 670 & \\

$v<700$kms$^{-1}$ ($R_i>20$)
      & $-20.01\pm0.01$ & $-0.93\pm0.01$ & 96 & 18 & 1470 & 600 & 96 \\
                  & $-20.56\pm0.07$ & $-1.25$ & 66 & & 800 &  \\
$v<700$kms$^{-1}$ ($R_i>15$)
      & $-19.98\pm0.04$ & $-0.92\pm0.03$ & 69 & 22 & 1700 & 660 & 88 \\
                  & $-20.56\pm0.08$ & $-1.25$ & 46 & & 900 &  \\

$v>700$kms$^{-1}$ ($R_i>20$)
      & $-20.31\pm0.02$ & $-1.66\pm0.01$ & 98 & 6 & 320 & 200 & 13\\
                  & $-19.66\pm0.07$ & $-1.25$ & 72 & & 840 & \\
$v>700$kms$^{-1}$ ($R_i>15$)
      & $-19.79\pm0.02$ & $-1.20\pm0.02$ & 88 & 7 & 730 & 220 & 88\\
                  & $-19.87\pm0.06$ & $-1.25$ & 95 & & 660 & \\

$10\le R_i<20$ & $-19.86\pm0.27$  & $-1.25\pm0.15$ & 7 & 15 & 700 & 220 & 69\\
                  & $-19.86\pm0.07$ & $-1.25$ & 13 & & 700 & \\
\end{tabular}
\caption{Fits to the subsets of the full cluster data as described in the text.
The columns are: the description of the subset; the derived values of $M^*$and
$\alpha$ (again, the error corresponds to the values of $M^*$ and $\alpha$
where $\chi^2 = \chi^2_{min}+1$); the formal likelihood that the fit is good;
the number of clusters that have been combined to form the composite for the
subset; the derived value of $n^*$; the measured value of the cluster richness;
lastly, the probability that the subset and the composite derived in section
4.1 have the same distribution using a 2-sample $\chi^2$ test.  The first row
for each entry gives the best two parameter fit, the second gives a one
parameter fit with $\alpha=-1.25$. Those entries that appear only in the first
row have the same values in the second. }
\end{table}

\clearpage

\noindent{\bf Figure Captions}\\

\noindent{\bf Figure 1:}  The relationship between the catalogued
EDSGC $b_j$ magnitude and the derived $b_j$ magnitude from the
CCD sequences.  The solid line represents the best fitting quadratic
given in the text.  The dashed line is given as a guide to the eye
to show the deviation between the CCD magnitude scale and the EDSGC
scale.

\noindent{\bf Figure 2:}  The number counts from the EDSGC derived
from the whole catalogue (the solid line), and from the average of all the
annuli used for background correction of the LF ($\bullet$).
The difference between the two never exceeds 10\%.  The error bars
shown represent the variance of the counts derived from the annuli.

\noindent{\bf Figure 3:} Derived values of $M^*$ from one parameter fits
to model Schechter luminosity functions for clusters with (a) $R_i=10$
and (b) $R_i=50$, when the fit is over the range $-18>M>-21$.  
The model LF has $M^*=-20.2$ and $\alpha=-1.25$.
When
this range is extended to $-18>M>-23$, the distribution of $M^*$
for $R_i=25$ is given in (c).  This is clearly similar to the higher richness
distribution when fitted over the narrower range in absolute magnitude.

\noindent{\bf Figure 4:} Contours of the derived distribution in $M^*$ and
$\alpha$ for 10000 simulations of a two parameter fit to the standard
Schechter function.  The initial model LF has $M^*=-20.2$ and $\alpha=-1.25$.
The contours contain 10\%, 68.5\%, 90\%, 95\% and 99\% of the data.  The 
initial richness given to the model LF is $R_c=250$, close to
the measured value for most of our subsamples in section 5.3.  This is
therefore the correct error map for comparison with the observed 
values of $M^*$ and $\alpha$ from those composite sub-samples.

\noindent{\bf Figure 5:}  Measured luminosity functions for 
(a) the local background correction method, (b) global background
correction and (c) global correction but using the average of the
annuli from (a).  The data are fitted to a Schechter function in the range
$-21<M<-18$.  The best fitting Schechter function is shown as a solid
line.  Only the data derived assuming a 50\% background variance
is shown.

\noindent{\bf Figure 6:} Measured luminosity functions for all those cluster
with measured richness values, $R_i$, of 15 or more using the local background
correction method.  The best one parameter fit to the 
 Schechter function is shown by a solid line.  The fit is over the range
$-21<M<M_{lim}$.

\noindent{\bf Figure 7:} Luminosity functions for (a) Bautz-Morgan class 1
clusters with $R_i>20$, (b) Bautz-Morgan class 3
clusters with $R_i>20$, (c) Bautz-Morgan class 1
clusters with $R_i>15$, (d) Bautz-Morgan class 3
clusters with $R_i>15$, (e) $R_i>20$ clusters,
(f) $10<R_i<20$ clusters,
(g) clusters with velocity dispersion $<700$kms$^{-1}$ and $R_i>20$,
(h) clusters with velocity dispersion $>700$kms$^{-1}$ and $R_i>20$,
(i) clusters with velocity dispersion $<700$kms$^{-1}$ and $R_i>15$,
(j) clusters with velocity dispersion $>700$kms$^{-1}$ and $R_i>15$.
The best fitting two parameter fits to a Schechter function over the
range $-18>M>-21$ are  shown as solid lines.  The best fitting one
parameter fits (with $\alpha=-1.25$) are shown as dashed lines.  As can be
seen,
there is often a marked difference between the best fitting one and two
parameter fits.

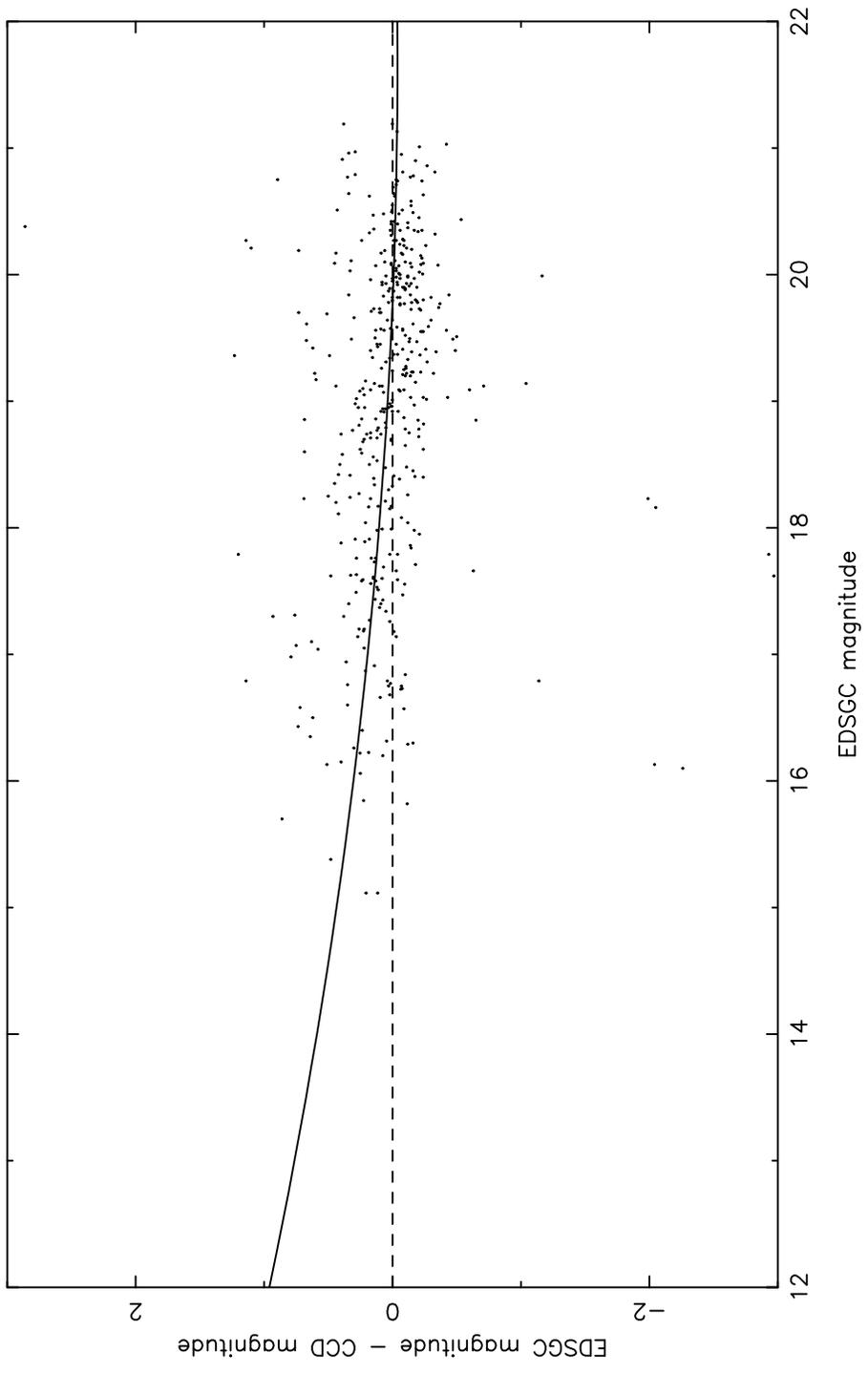
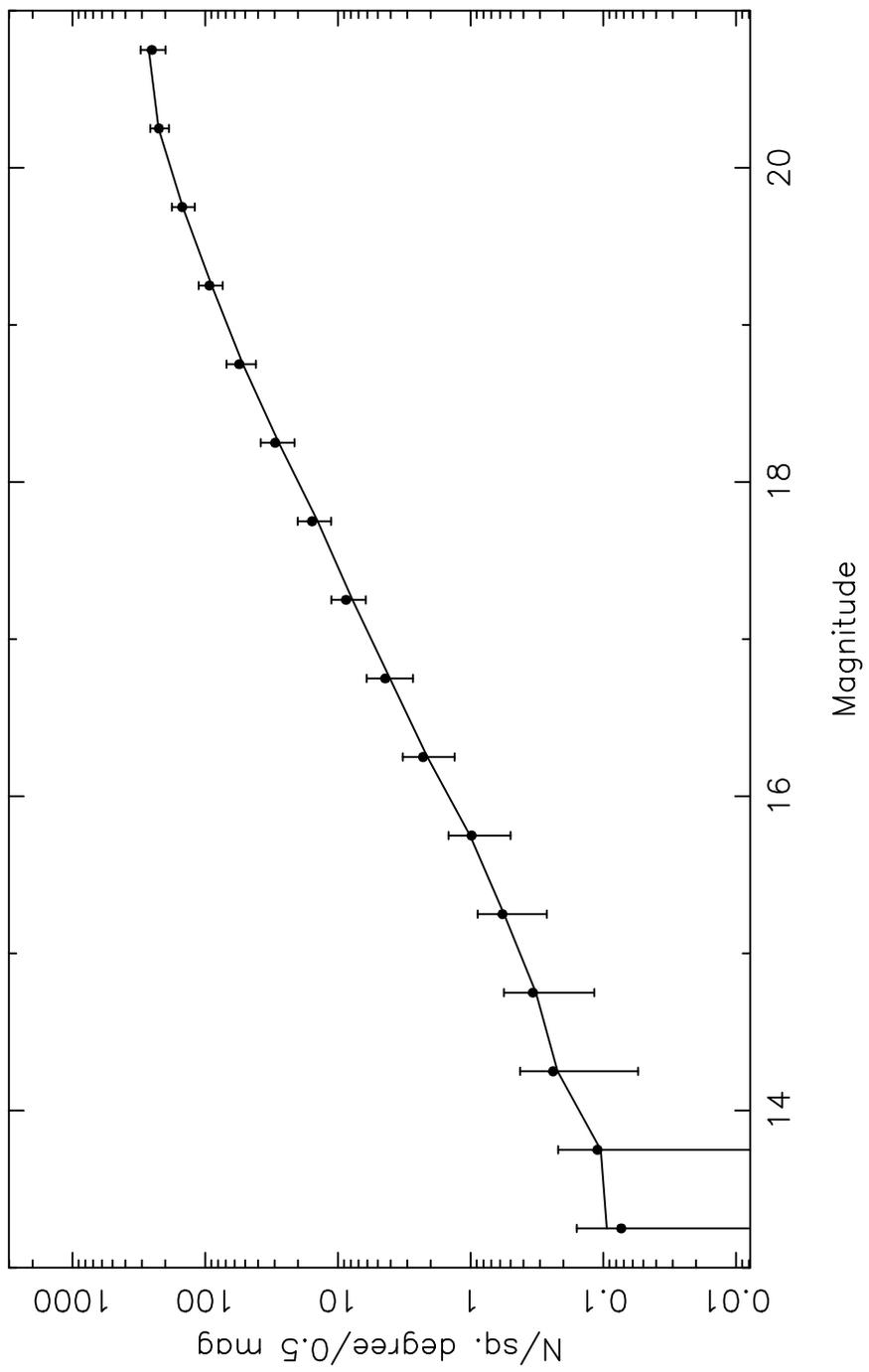
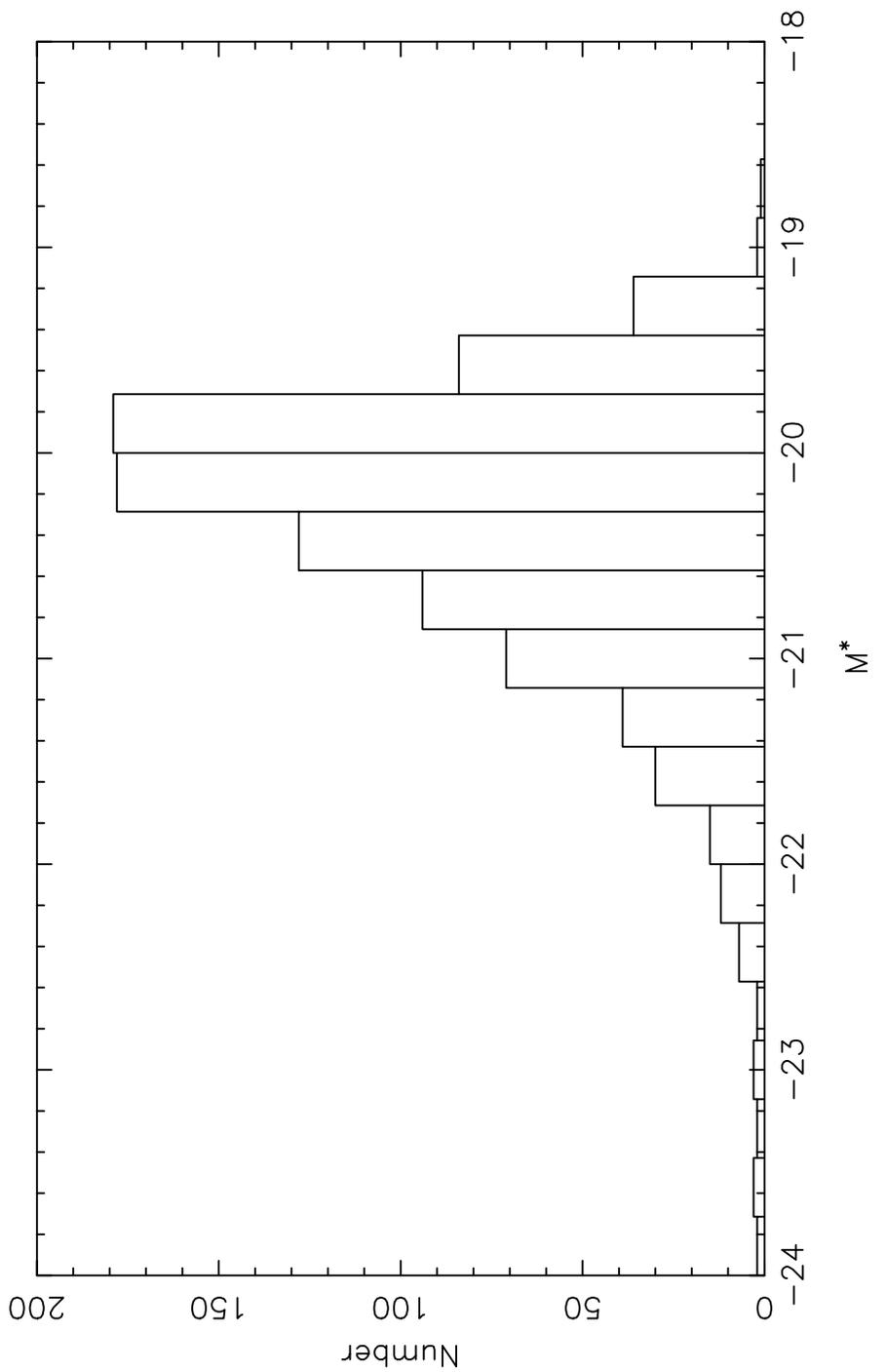
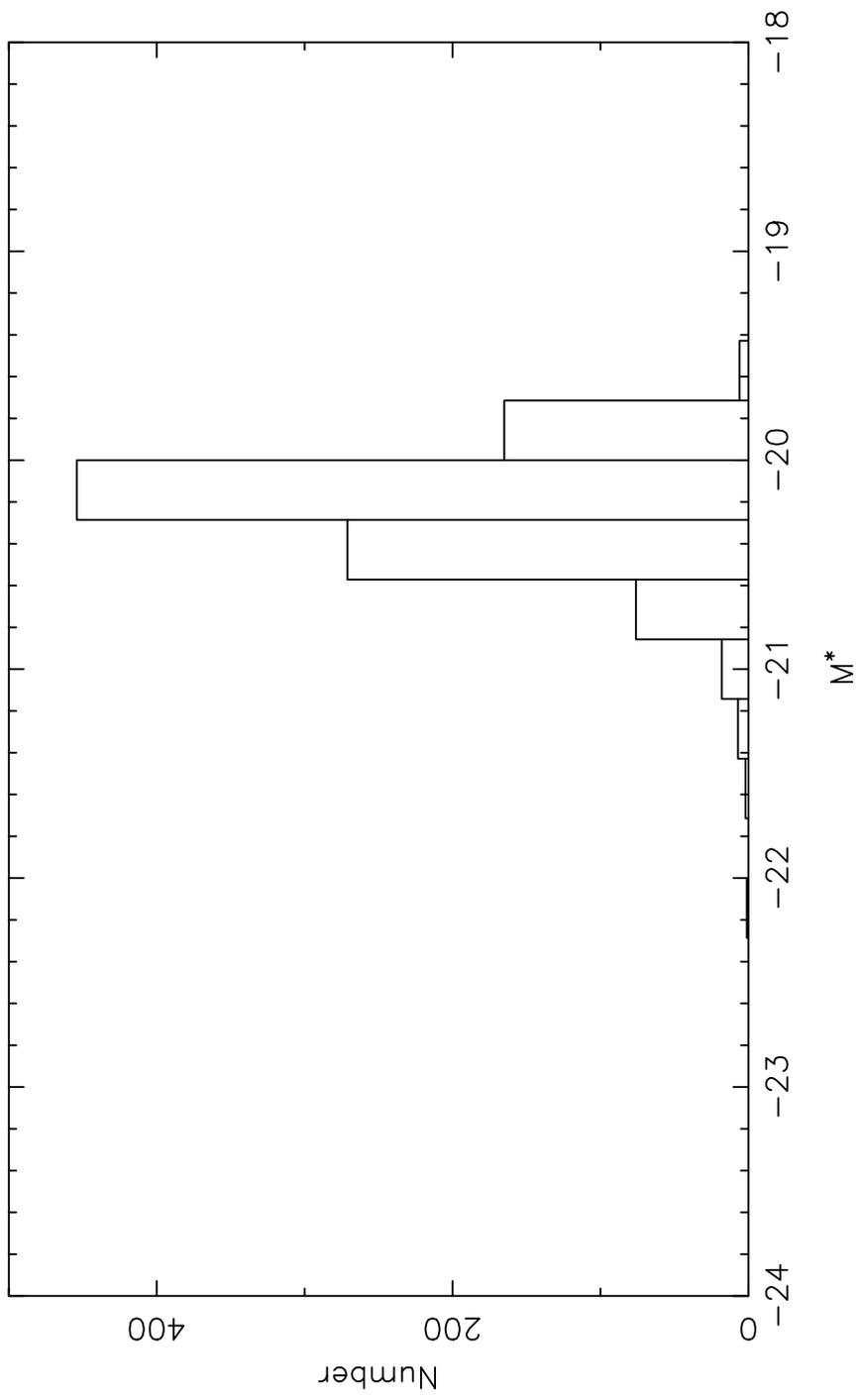
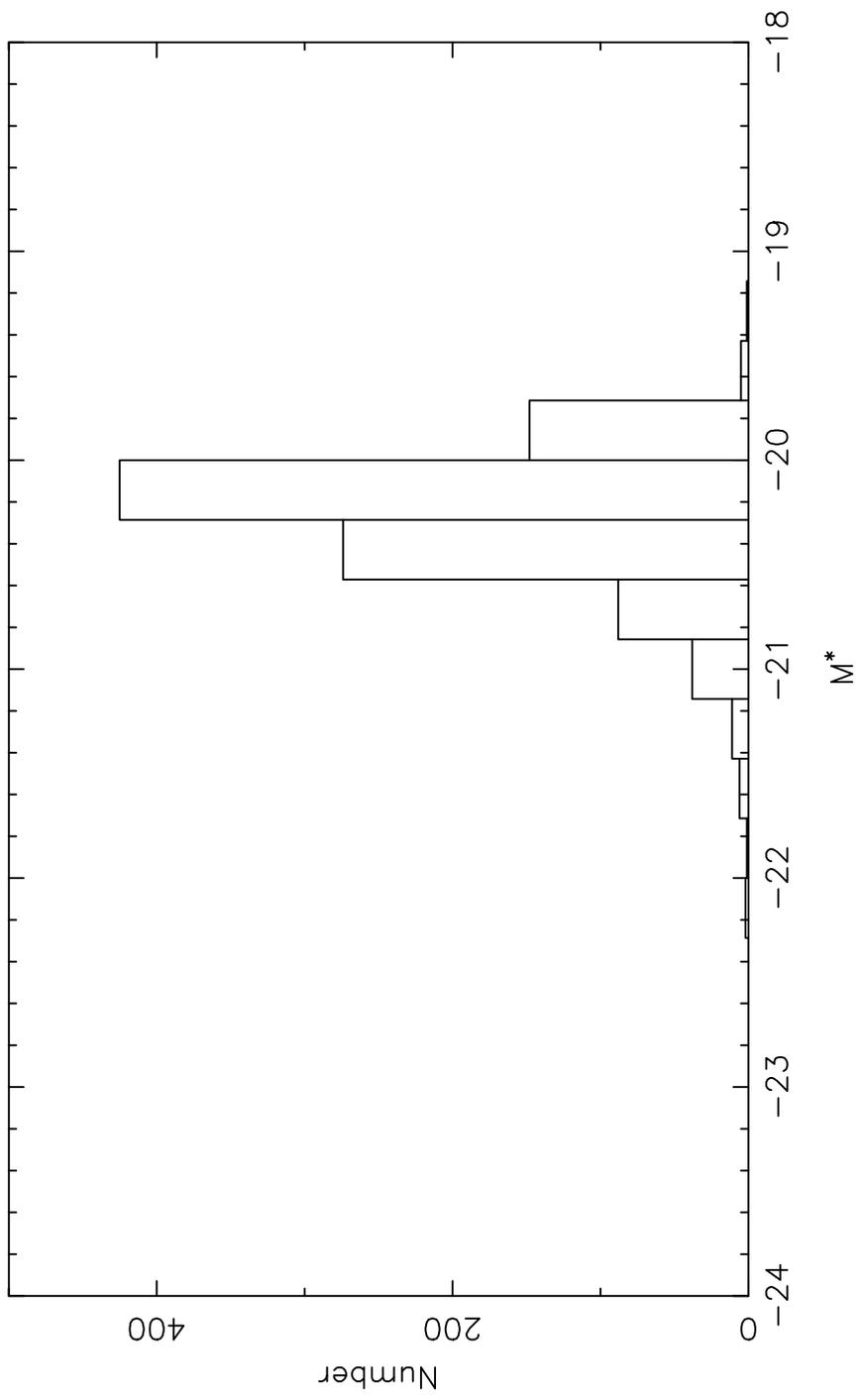
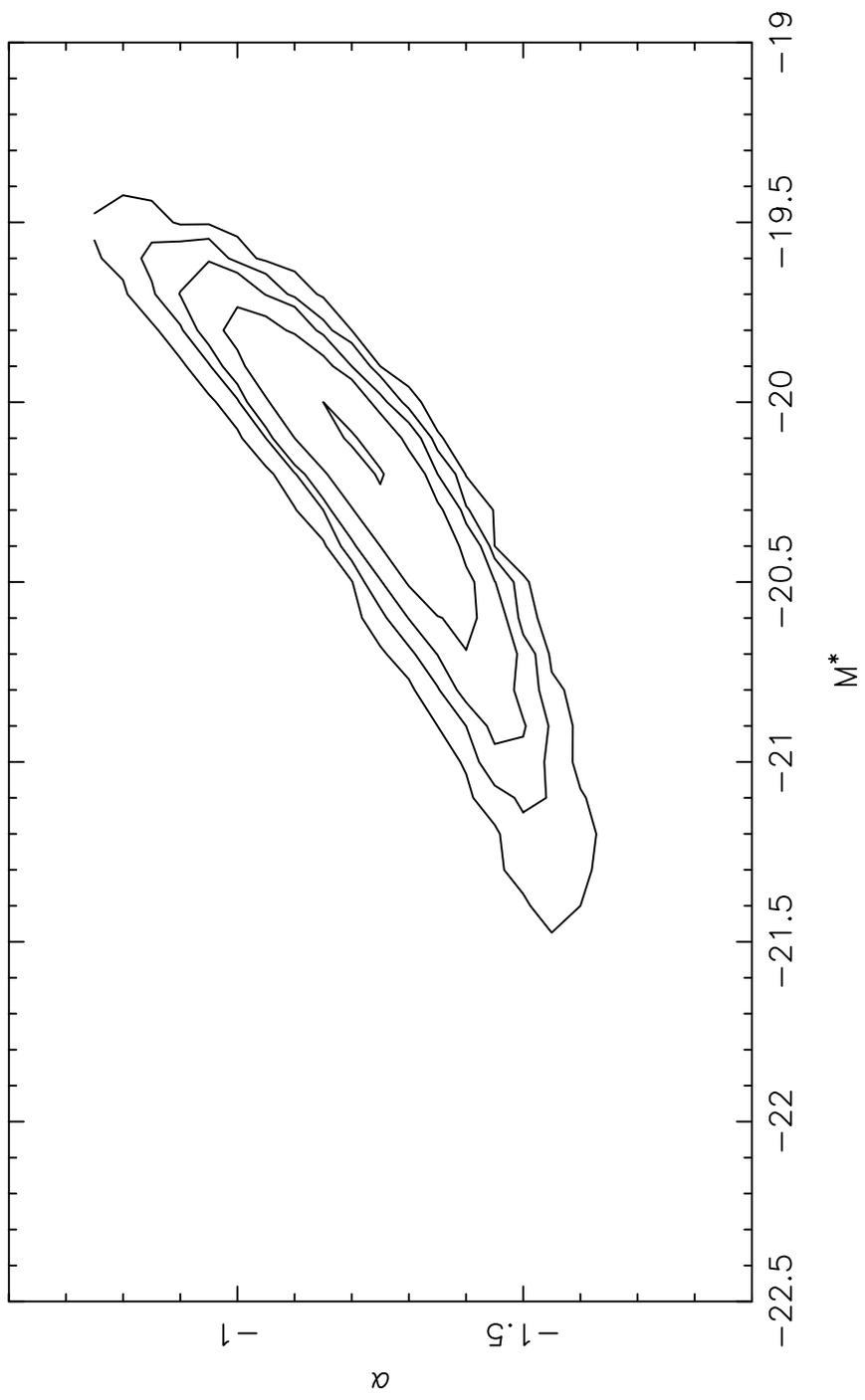
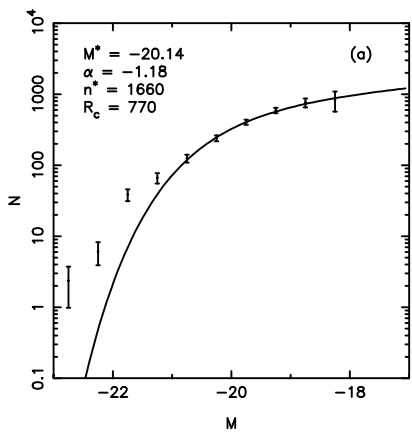
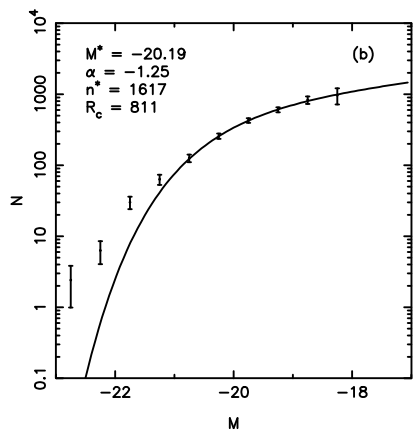
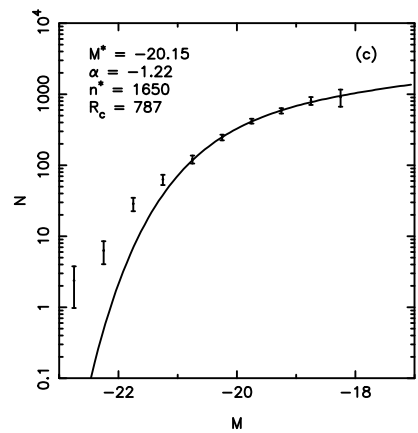
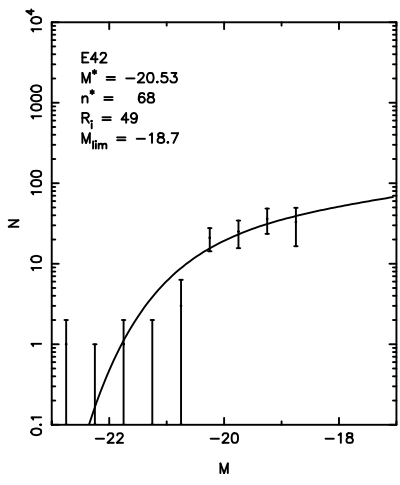
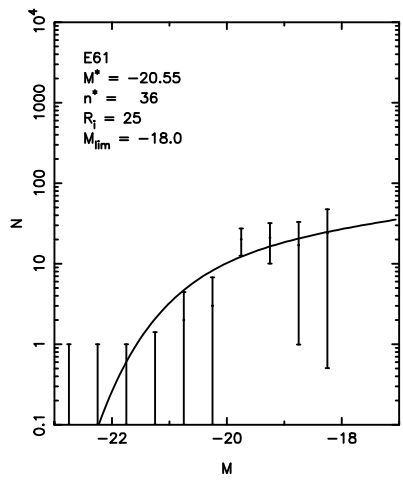
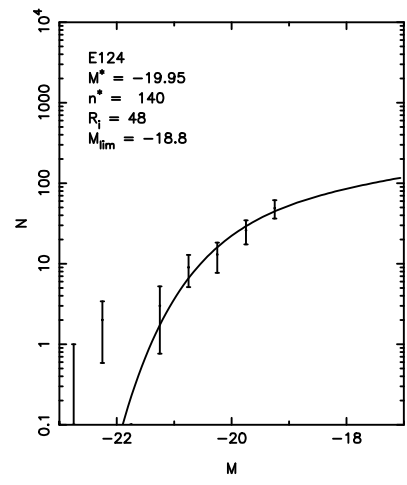
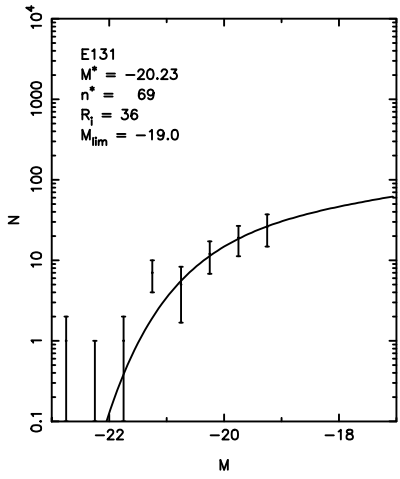
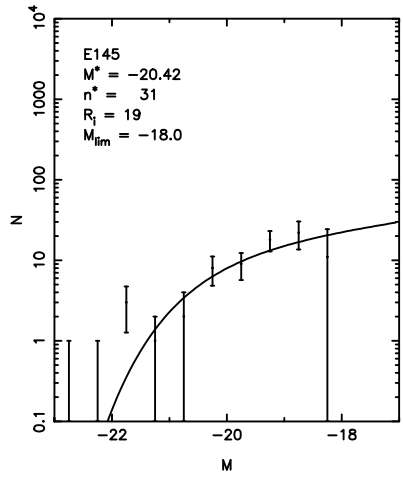
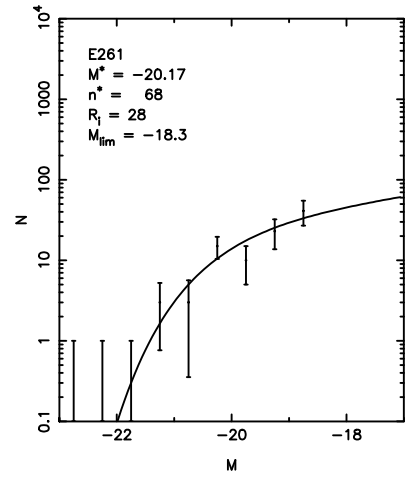
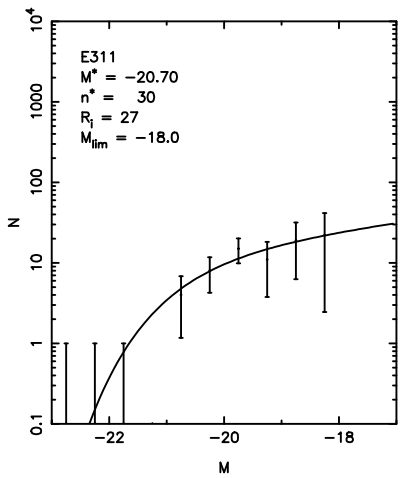
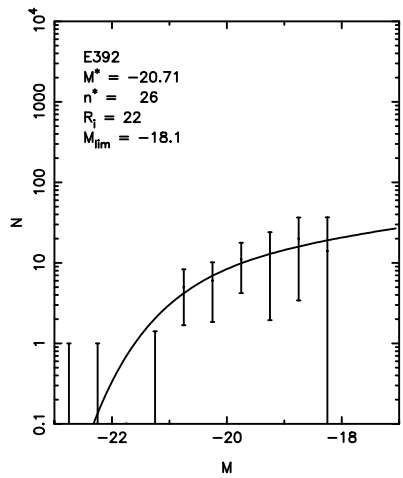
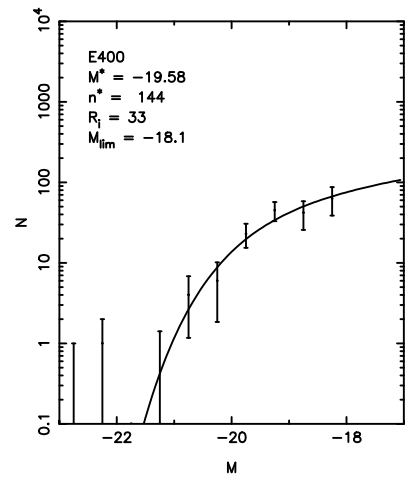
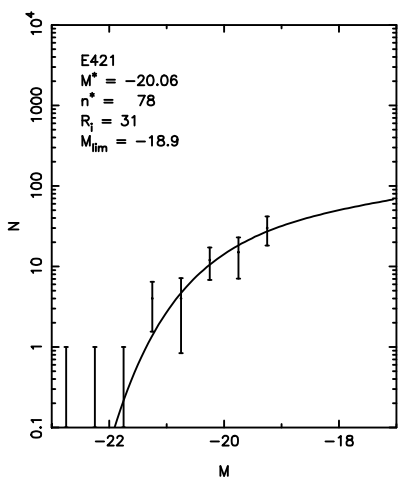
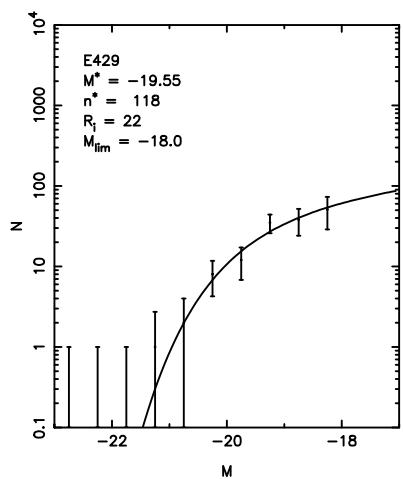
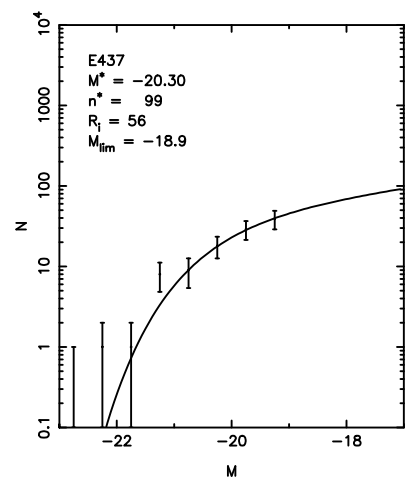
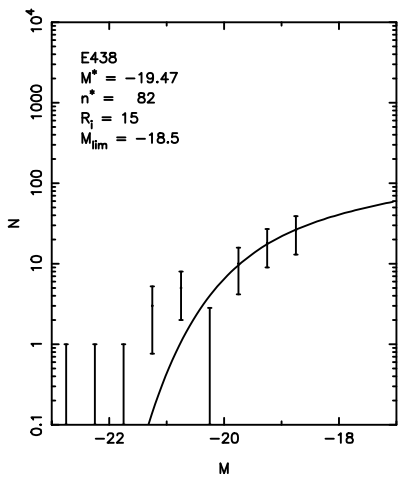
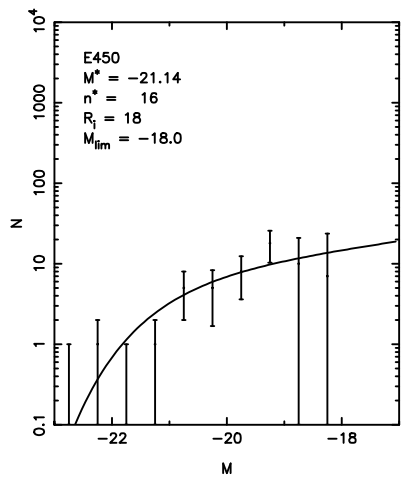
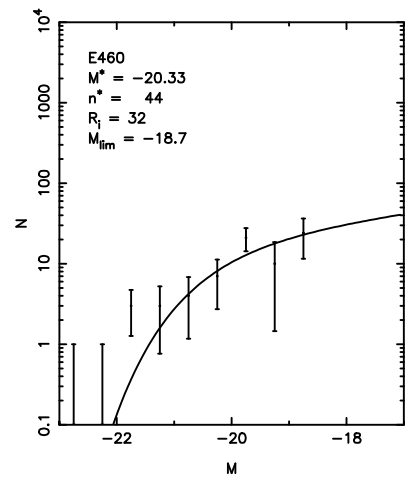
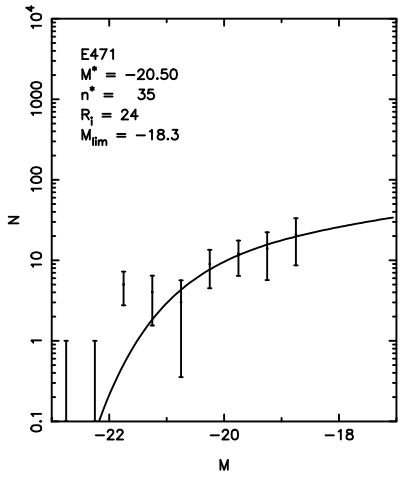
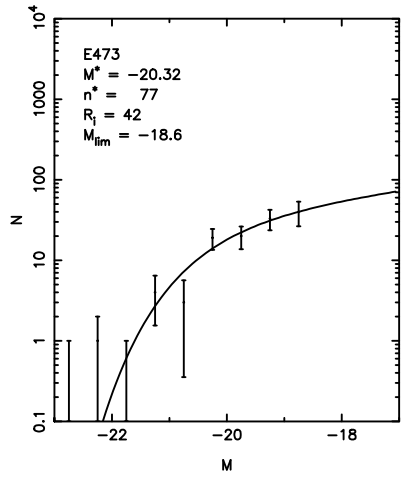
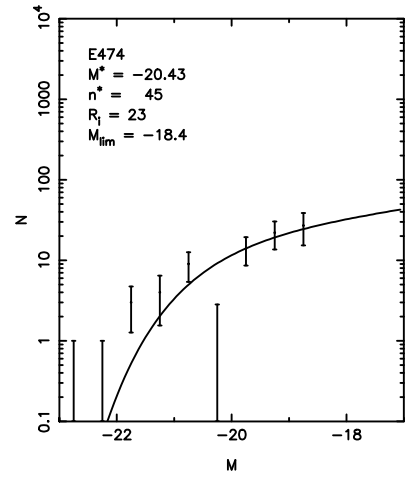
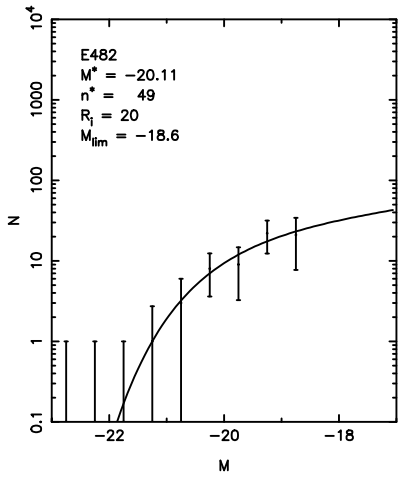
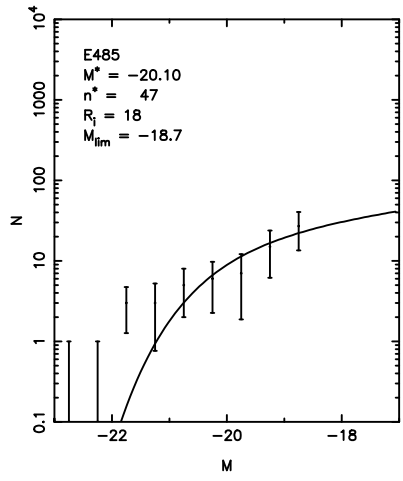
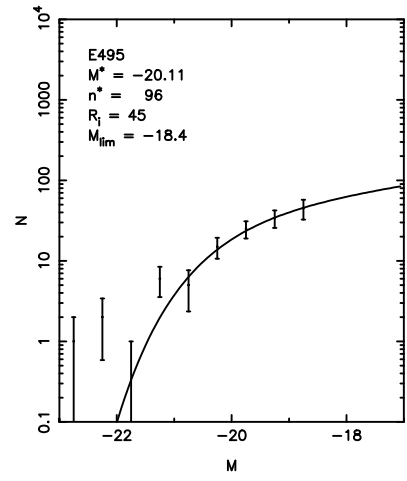
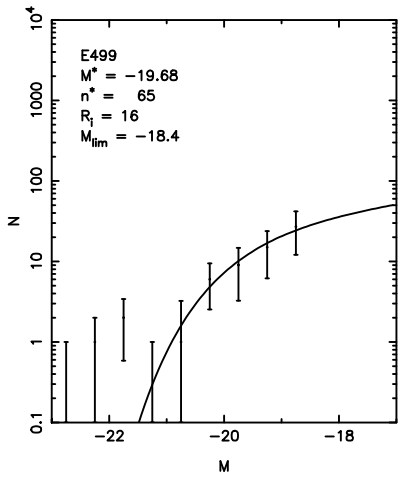
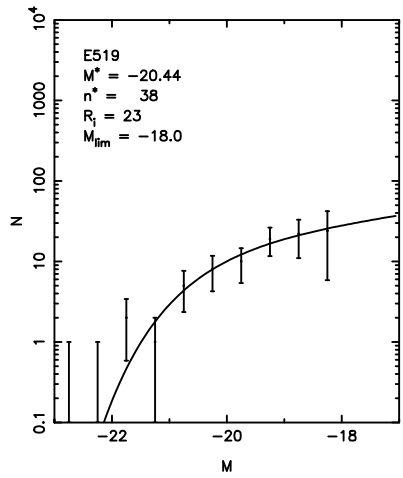
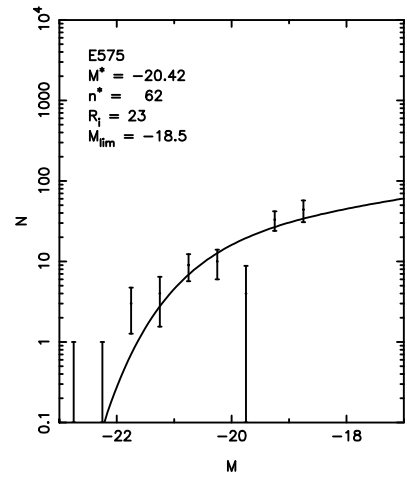
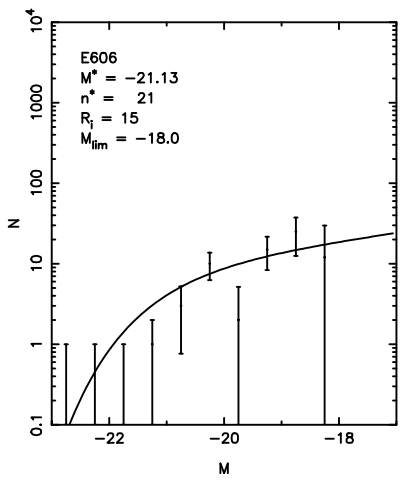
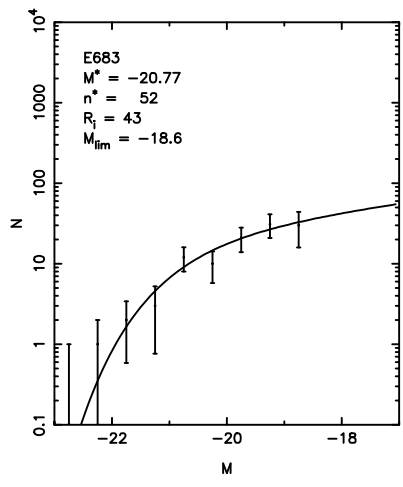
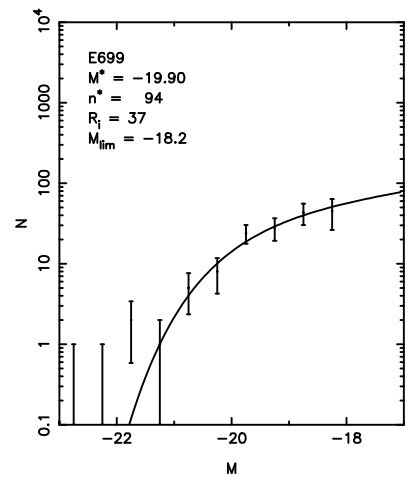
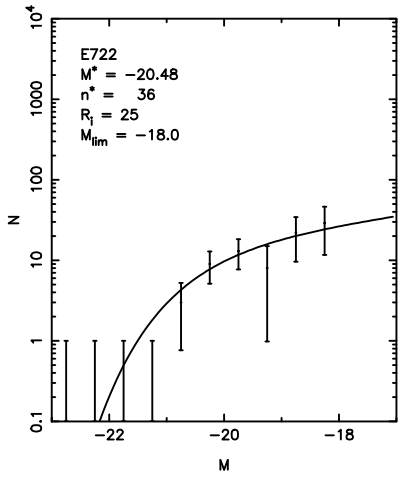
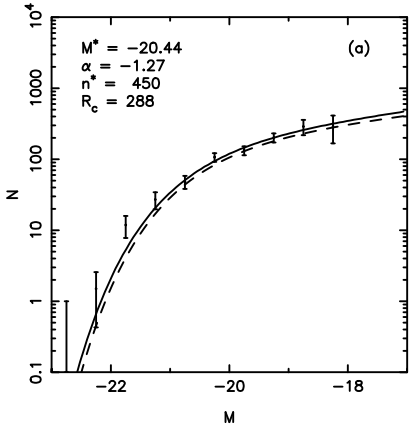
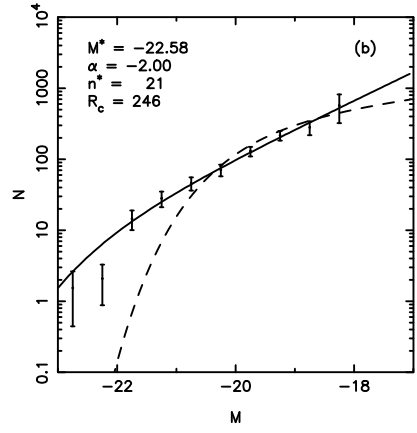
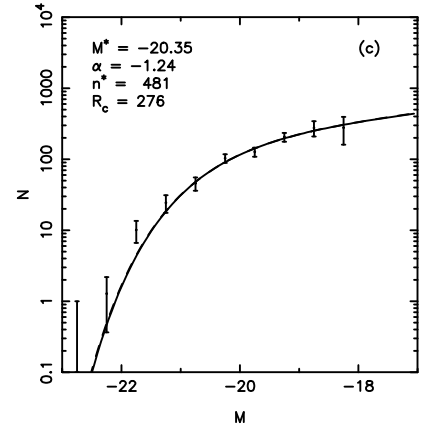
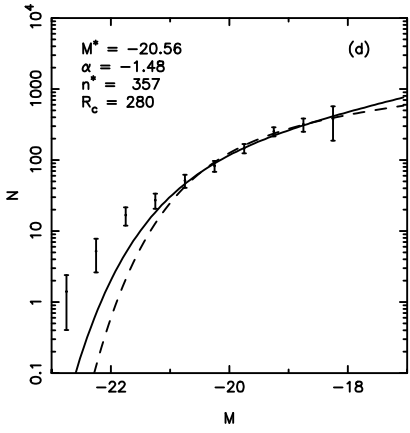
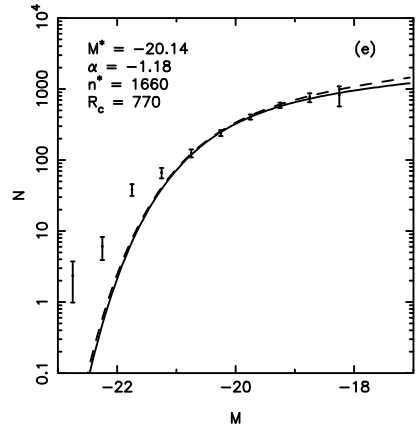
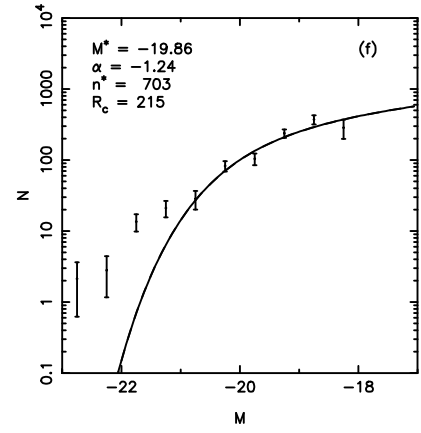
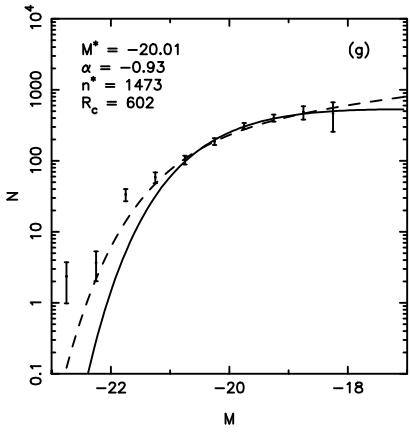
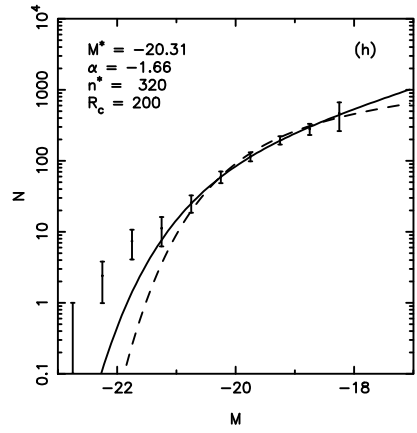
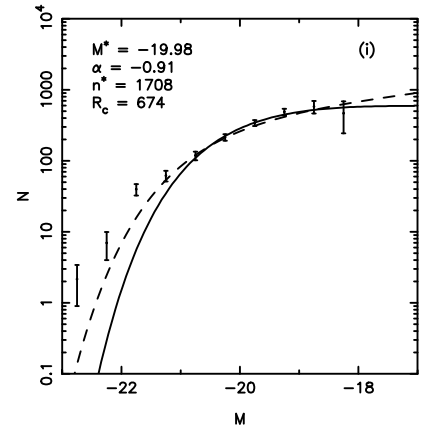
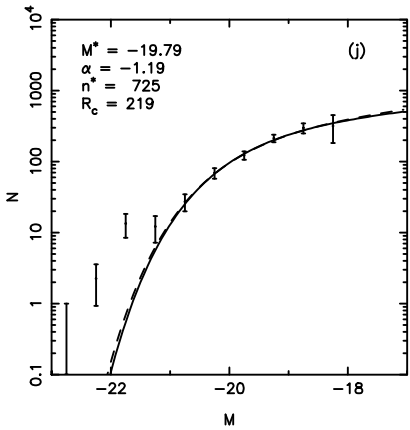
\end{document}